\documentclass[prb,aps,twocolumn,groupedaddress,floatfix,citeautoscript,showpacs]{revtex4}
\usepackage{graphicx,rotating,subfigure,amsmath,amsfonts,amssymb,delarray}
\usepackage{color}
\usepackage{dsfont}
\usepackage[T1]{fontenc}

\newcommand{\e}{\text{e}}

\def\12{\frac{1}{2}}

\newcommand{\be}{\begin{equation}}
\newcommand{\ee}{\end{equation}}
\newcommand{\bea}{\begin{eqnarray}}
\newcommand{\eea}{\end{eqnarray}}

\DeclareMathOperator\Tr{Tr}

\begin{document}
\bibliographystyle{apsrev} 

\title{Entanglement entropy of disordered quantum chains following a
  global quench}

\author{Y. Zhao}
\author{F. Andraschko}
\author{J. Sirker}
\affiliation{Department of Physics and
  Astronomy, University of Manitoba, Winnipeg R3T 2N2, Canada}

\date{\today}

\begin{abstract}
  We numerically investigate the growth of the entanglement entropy
  $S_{\rm ent}(t)$ in time $t$---after a global quench from a product
  state---in quantum chains with various kinds of disorder. The main
  focus is, in particular, on fermionic chains with bond disorder. In
  the noninteracting case at criticality we numerically test recent
  predictions by the real space renormalization group for the
  entanglement growth in time, the maximal entanglement as a function
  of block size, and the decay of a density wave order parameter. We
  show that multiprecision calculations are required to reach the
  scaling regime and perform such calculations for specific cases. For
  interacting models with binary bond disorder we present data based
  on infinite size density matrix renormalization group calculations
  and exact diagonalizations. We obtain first numerical evidence for a
  many-body localized phase in bond disordered systems where $S_{\rm
  ent}(t)\sim\ln t$ seems to hold. Our results for bond disorder are
  contrasted with the well studied case of potential disorder.
\end{abstract}

\pacs{75.10.Jm, 05.70.Ln, 72.15.Rn}

\maketitle

\section{Introduction}
A method commonly used to study nonequilibrium dynamics in cold atomic
gases and trapped ion systems is a global quench, a sudden change of a
global control
parameter.\cite{GreinerMandel,BlochDalibard,UlmRossnagel} Numerically,
such dynamics can be studied by approximating the time evolved quantum
state by a matrix product state
(MPS).\cite{VidalTEBD1,VidalTEBD2,DaleyKollath,WhiteFeiguin,SirkerKluemperDTMRG,EnssSirker,Schollwock_MPS_review}
The time interval accessible by this method is determined by the rate
at which the entanglement grows because an MPS with a finite matrix
dimension can only faithfully approximate weakly entangled states. It
is therefore of great interest to understand precisely how the
entanglement depends on time following a quantum quench. While this
question has already been studied extensively for clean
systems\cite{CalabreseCardyQuench,CalabreseCardy2005,Laflorencie_review},
disordered systems have only recently come into
focus.\cite{ZnidaricProsen,IgloiSzatmari,BardarsonPollmann,AndraschkoEnssSirker} In
such systems entanglement can, in addition, also provide a novel
viewpoint to study localization. This is particularly useful in
interacting many-body systems where a picture based on single particle
states and related localization measures are not applicable.

In a generic clean quantum system we expect that the entanglement of
most eigenstates follows a volume law. A well-known exception is the
ground state of a local Hamiltonian which typically shows an area law
scaling.\cite{HolzheyLarsen} Given that a generic initial state in a quantum
quench is a linear combination of many different eigenstates of the
Hamiltonian responsible for the unitary time evolution, the
entanglement of an infinite system, in general, grows without bounds
as a function of time. A commonly used measure of entanglement for
many-body systems is the entanglement entropy
\begin{equation}
S_{\textrm{ent}}(t) =-\Tr \rho_A(t) \ln \rho_A(t)
\end{equation}
which is the von-Neumann entropy of a reduced density matrix
$\rho_A(t)=\Tr_B\rho(t)$ obtained from the regular density matrix
$\rho(t)$ by splitting the system into two parts $A$ and $B$ and
taking a partial trace.

The entanglement growth $S_{\textrm{ent}}(t)$ is best understood for a
global quench in a one-dimensional conformally invariant
system.\cite{CalabreseCardy2005,CalabreseCardy09} In this case,
conformal field theory predicts a linear increase of the entanglement
in time for $vt<\ell$ where $v$ is the velocity of excitations and
$\ell$ the length of block $A$. For $vt>\ell$ the entanglement entropy
saturates. This behavior can physically be understood in a picture of
entangled quasiparticles which move in opposite
directions.\cite{CalabreseCardy09} The lightcone-like spreading
implied by this picture is consistent with the Lieb-Robinson
bounds.\cite{LiebRobinson,BravyiHastings}

Conformal field theory works surprisingly well even on a quantitative
level for quenches in one-dimensional critical lattice models where
conformal invariance only holds approximately at low energies.
Consider, for example, the free spinless fermion model (XX model) with
Hamiltonian
\begin{equation}
\label{XX}
H=-\sum_i J_i (c_i^\dagger c_{i+1} + h.c.) -\sum_i \mu_i (c_i^\dagger c_i -1/2)
\end{equation}
in the clean case $J_i=J\equiv 1$ and with $\mu_i=0$. Here
$c_i^{(\dagger)}$ annihilates (creates) a spinless fermion at site
$i$. As initial state for the quantum quench we consider in the
following---unless stated otherwise---the density wave product state
\begin{equation}
\label{Neel}
|D\rangle = \prod_i c^\dagger_{2i} |0\rangle
\end{equation}
where $|0\rangle$ denotes the vacuum. At low energies, the spectrum
can be linearized around the two Fermi points and the model becomes
conformally invariant in this approximation. 

\begin{figure}[ht!]
\includegraphics*[width=0.99\columnwidth]{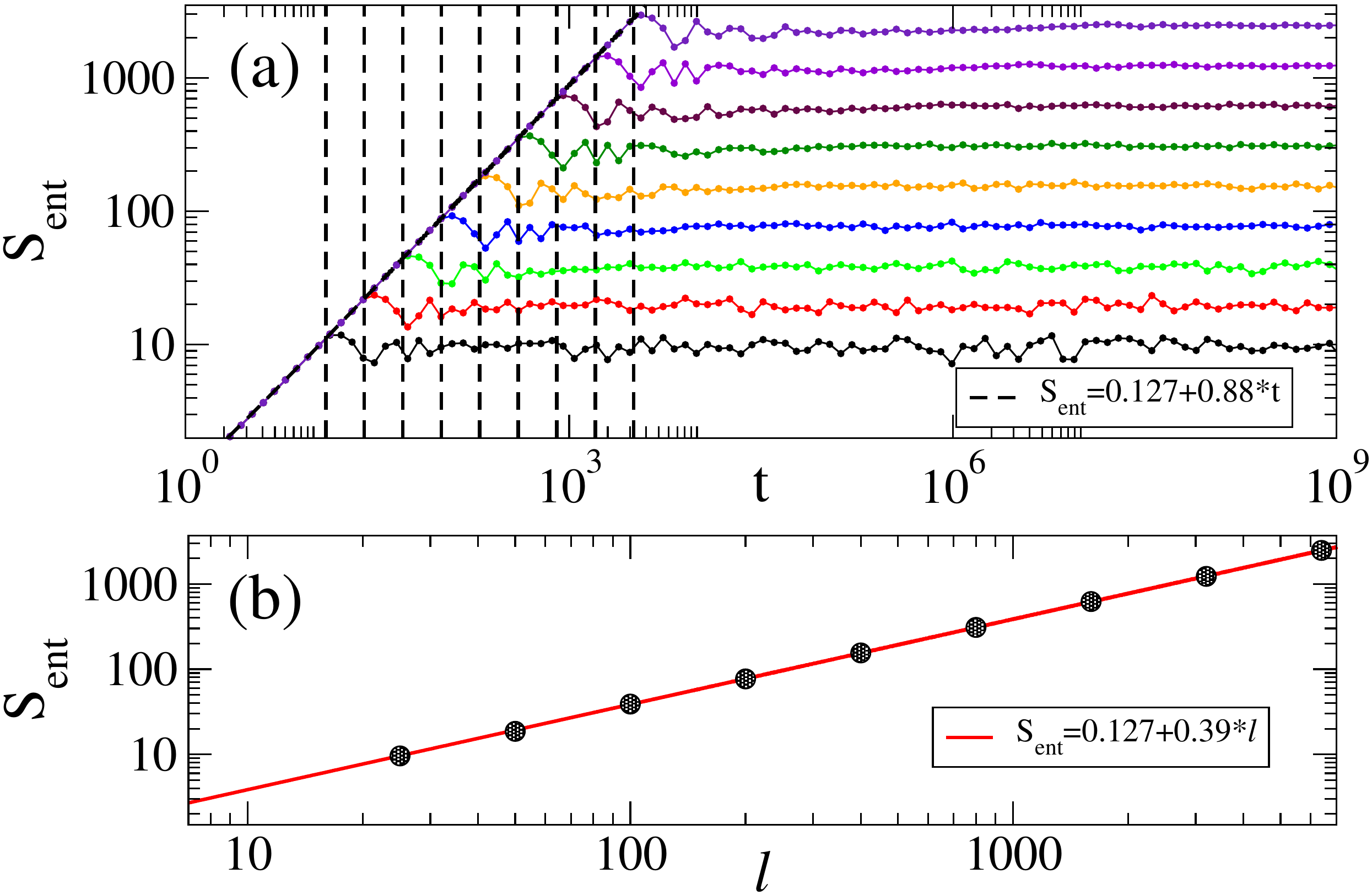}
\caption{(Color online) (a) $S_{\rm ent}(t)$ for the open XX chain
  without disorder. Symbols denote ED data for system sizes
  $N=50,100,200,400,800,1600,3200,6400, 12800$ (from bottom to top),
  solid lines are a guide to the eye.  The dashed vertical lines
  denote the crossover scale $t^*=\ell/2$. (b) Symbols: Saturation
  value as a function of block size $\ell$. The fits in (a) and (b)
  show that $S_{\rm ent}\sim at$ for $t<t^*$ and a saturation at
  $S_{\rm ent}\sim a/2\cdot \ell$ for $t>t^*$.}
\label{Fig1}
\end{figure}

For a free fermion model such as the XX model in Eq.~\eqref{XX} the
eigenvalues of the reduced density matrix $\rho_A(t)$ can be obtained
by an exact diagonalization (ED) of the matrix of two-point
correlations in the time evolved state $|D(t)\rangle
=\e^{-iHt}|D\rangle$, see 
Refs.~\onlinecite{ChungPeschel,Peschel2004,PeschelEisler} for details.
In Fig.~\ref{Fig1}, results for $S_{\textrm{ent}}(t)$ obtained
numerically for model \eqref{XX} and various system sizes $N$ are
shown. Here we have used open boundary conditions (OBC) and the block
size is fixed to $\ell=N/2$.
In this case, conformal field theory predicts that 
\begin{equation}
\label{CFT}
S_{\rm ent} \sim \left\{\begin{array}{l} \frac{\pi ct}{12\tau_0} \quad vt<\ell \\*[2pt] \frac{\pi c\ell}{24\tau_0}\quad vt>\ell \end{array}\right.
\end{equation}
where $v=2$ with central charge $c=1$.\cite{CalabreseCardy09} We
therefore expect a crossover at times $t^*=\ell/v=N/4$ from a linear
increase to saturation. This behavior is nicely confirmed by the data
shown in Fig.~\ref{Fig1}(a). Note also that while the scale $\tau_0$
in Eq.~\eqref{CFT} is nonuniversal, the ratio of the slope of the
linear increase at times $t<t^*$ and the saturation value at $t>t^*$
is universal because $\tau_0$ cancels out. This quantitative
prediction is also approximately fulfilled for the considered lattice
model as can be seen from the fits in Fig.~\ref{Fig1}(a) and (b).

Recently, there is a renewed interest in localization phenomena in
{\it interacting} many-body
systems.\cite{BaskoAleiner,NandkishoreHuse,AltmanVoskReview} In
one-dimensional Heisenberg models with potential disorder it has been
argued---based on exact diagonalization data---that there is a
transition at finite disorder strength between an ergodic and a
many-body localized (MBL) phase.\cite{PalHuse} One of the main
features used to identify an MBL phase after a quantum quench is the
slow logarithmic growth of the entanglement entropy
\cite{BardarsonPollmann}. This has to be contrasted with the linear
growth in the clean case and a saturation,
$\lim_{t\to\infty}\lim_{\ell\to\infty}\lim_{N\to\infty}S_{\textrm{ent}}(N,\ell,t)=\mbox{const}$,
which might naively be extected in an Anderson insulating phase of
noninteracting particles due to the finite localization length.

While disorder is always a relevant perturbation for a non-interacting
one-dimensional quantum system,\cite{AbrahamsAnderson} critical points
or lines between localized phases can still exist and the above
picture for the entanglement growth in an Anderson insulator is, in
general, too naive. Two well-known examples where critical behavior in
a disordered system occurs are the transverse Ising chain and the XX
model with bond disorder.
\cite{EggarterRiedinger,Fisher94,Fisher_random_Ising,Fisher_random_Ising2,BalentsFisher97}
A real-space renormalization group (RSRG) treatment---where the
strongest bonds are successively eliminated---predicts that such
systems are driven to an infinite randomness fixed point where the
{\it mean localization length} scales as
\begin{equation}
\label{scaling}
\xi_{\rm loc}(\varepsilon) \sim |\ln(\varepsilon)|^\Psi
\end{equation}
with a critical exponent $\Psi$ and is much larger than the {\it
  typical localization length}.\cite{BalentsFisher97} These systems
therefore show a delocalization transition as a function of
eigenenergy $\varepsilon$.  For the ground state entanglement entropy
of a block of length $\ell$ in an infinite critical random system this
implies $S_{\rm ent}\sim\ln\ell$.\cite{RefaelMoore} For the XX model
with bond disorder the latter prediction has been confirmed by an
extensive numerical study.\cite{Laflorencie05}

Lately, the RSRG treatment of critical disordered systems has been
generalized to excited states by also allowing for projections of a
bond onto higher energy
states.\cite{RefaelPekker,VoskAltman,VoskAltman2} This so-called
RSRG-X approach allows, in particular, to make predictions with regard
to the time evolution of the entanglement following a quantum quench
from a product state in non-interacting critical random systems. Since
RSRG-X essentially yields the same infinite randomness structure for
excited states as for the ground state, a logarithmic scaling with
block size $\ell$
\begin{equation}
\label{saturation}
\lim_{t\to\infty}\lim_{N\to\infty}S_{\rm ent} \sim b\ln\ell
\end{equation}
of the saturation value is expected, mimicking the ground state
behavior. In the infinite system at large finite times we therefore
expect $S_{\rm ent}\sim \ln L(t)$ where $L(t)\leq\ell$ is a length
scale which, according to Eq.~\eqref{scaling}, is expected to scale as
$L(t)\sim |\ln t|^\Psi$.  This implies
\begin{equation}
\label{t-dep}
\lim_{\ell\to\infty}\lim_{N\to\infty}S_{\rm ent} \sim a\ln(\ln t)
\end{equation}
in the infinite system at long times. Furthermore, the ratio of the
prefactors is given by $a/b=\Psi$ with the critical exponent $\Psi$
defined in Eq.~\eqref{scaling}. Numerically, a $\ln(\ln t)$ scaling in
a certain time interval has been observed for the critical random
transverse Ising model.\cite{IgloiSzatmari} Another specific
prediction of the RSRG concerns a quench in the XX model from an
initial state with density wave order, Eq.~\eqref{Neel}. In this case,
the density wave order parameter is expected to decay as $\Delta n\sim
1/\ln^2 t$ at long times.\cite{VoskAltman} The latter result is
expected to hold also in interacting models which are in a critical
MBL phase.  Eq.~\eqref{t-dep}, however, is expected to change to
$S_{\rm ent}\sim (\ln t)^\alpha$ with $\alpha\geq 1$ in the presence of
interactions. Furthermore, the saturation value,
Eq.~\eqref{saturation}, will then follow a volume law, $S_{\rm
ent}\sim\ell$, but will remain smaller than in a system where the
subsystem $A$ reaches thermal equilibrium.

In this paper we will present a thorough numerical analysis of the
entanglement dynamics in generic and critical one-dimensional systems
which can be described in terms of free fermions. In particular, we
will test the various scaling predictions above which have been
obtained based on the RSRG-X approach. Our work extends and
generalizes previous numerical studies of the entanglement dynamics in
such systems \cite{ChiaraMontangero06,IgloiSzatmari} by considering
both XX and transverse field Ising models with various types of
disorder and by providing numerical data for larger systems, larger
sample sizes, and including multiprecision data required to access
long times. For the interacting XXZ model we will try to bridge the
gap between RSRG-X predictions for bond disorder quenching from the
N'eel state and numerical studies which have so far concentrated
mostly on potential disorder. Checking the RSRG-X predictions for the
bond disordered case numerically is important because the RSRG-X
results for critical MBL phases are not applicable to the case of
potential disorder.

Our paper is organized as follows: As an example for a generic
disordered system of non-interacting particles we study
$S_{\textrm{ent}}(t)$ for the XX model, Eq.~\eqref{XX}, with potential
disorder in section \ref{Pot_disorder}. In section \ref{Bond_disorder}
we present results for the critical XX model with bond disorder. In
both sections we consider a box as well as a binary disorder
distribution. 
In section \ref{Ising} we show that qualitatively similar results as
for the critical XX model with bond disorder are also obtained for the
critical random transverse Ising model. Using exact diagonalizations
of small systems and density-matrix renormalization group (DMRG)
calculations for infinite systems with binary bond disorder we then
investigate in section \ref{Interaction} the entanglement growth once
interactions are included. In the final section \ref{Conclusions} we
summarize our main results and discuss them in the light of the RSRG-X
predictions and recent numerical studies of MBL phases.

\section{Generic disordered chains of noninteracting fermions}
\label{Pot_disorder}
As an example for a generic, non-critical disordered system we
consider in the following the XX chain, Eq.~\eqref{XX}, with potential
disorder $\mu_i$ drawn either from a box or a binary distribution, and
$J_i\equiv J=1$ fixed.

Without disorder, there are three length scales in the problem: the
chain length $N$, the dynamical scale $vt$, and the block length
$\ell$. For most parts of the paper we keep the block length fixed,
$\ell=N/2$, reducing the number of independent length scales to $2$.
In the clean case there are therefore only two regimes with different
scaling, see Fig.~\ref{Fig1}. With disorder, on the other hand, we
have instead three independent length scales: $L_{\textrm{ent}}(t)$,
$\ell=N/2$, and $\xi_{\rm loc}$. Here $L_{\textrm{ent}}(t)$ denotes
the length scale over which particles are entangled and $\xi_{\rm
loc}$ is the localization length which, in general, will depend on
energy $\xi_{\rm{loc}}=\xi_{\rm{loc}}(\varepsilon)$. The simplest case
is $\xi_{\rm loc}\gg \ell\gg 1$ corresponding to weak disorder in
which case the new length scale $\xi_{\rm loc}$ is completely
irrelevant for the dynamics and we expect $L_{\textrm{ent}}(t)\sim vt$
so that the results for the case without disorder approximately
hold. The entanglement properties are expected to change once
$\ell\sim\xi_{\rm loc}$. It is this regime which we want to analyze in
detail in the following.

\subsection{Box distribution}
First, we consider box potential disorder $\mu_i\in [-W/2,W/2]$. We
average over at least 2000 samples and make sure that the data are
converged by comparing results where a different number of samples
have been kept.

\begin{figure}[ht!]
\includegraphics*[width=0.99\columnwidth]{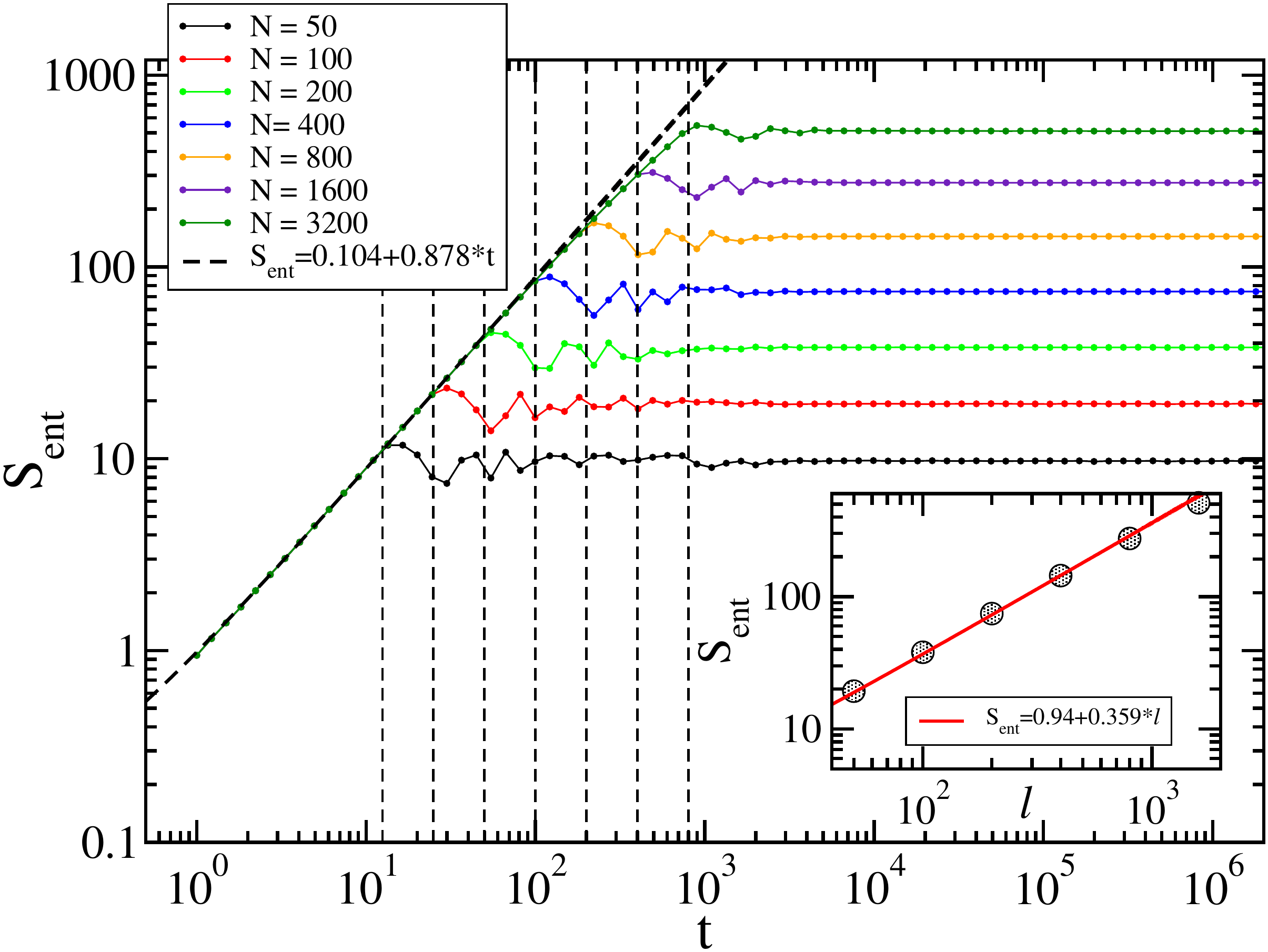}
\caption{(Color online) $S_{\rm ent}$ for the open XX chain with box
  potential disorder $W=0.1$. The dashed black lines denote the
  crossover scale $t^*=\ell/2$ in the clean case. Inset: Saturation
  value $S_{\rm ent}(\ell)$ for $t\to\infty$. The fits show that
  $S_{\rm ent}\sim at$ for $t<t^*$ and $S_{\rm ent}\sim a/2\cdot \ell$
  for $t>t^*$ still approximately hold if $N<3200$.}
\label{Fig2}
\end{figure}

For very weak disorder, $W=0.1$, and chain lengths up to $N=1600$, we
are in a regime with $\xi_{\rm loc}\gg \ell$ where the entanglement
growth is indeed very similar to the case without disorder, see
Fig.~\ref{Fig2}.
Only for the largest length shown, $N=3200$, do deviations from a
linear increase and a crossover at $t^*=N/4$ become clearly visible.
For the localization length this means that $\xi_{\rm loc}\gtrsim
3200$. The increase of the entanglement entropy in time and the
saturation value as a function of block length $\ell$ are still
approximately linear for lengths $N<3200$, see the inset of
Fig.~\ref{Fig2}. Note, however, that already for small system sizes
the fluctuations in the saturation value are noticeably surpressed
compared to the clean case.

\begin{figure}[ht!]
\includegraphics*[width=0.99\columnwidth]{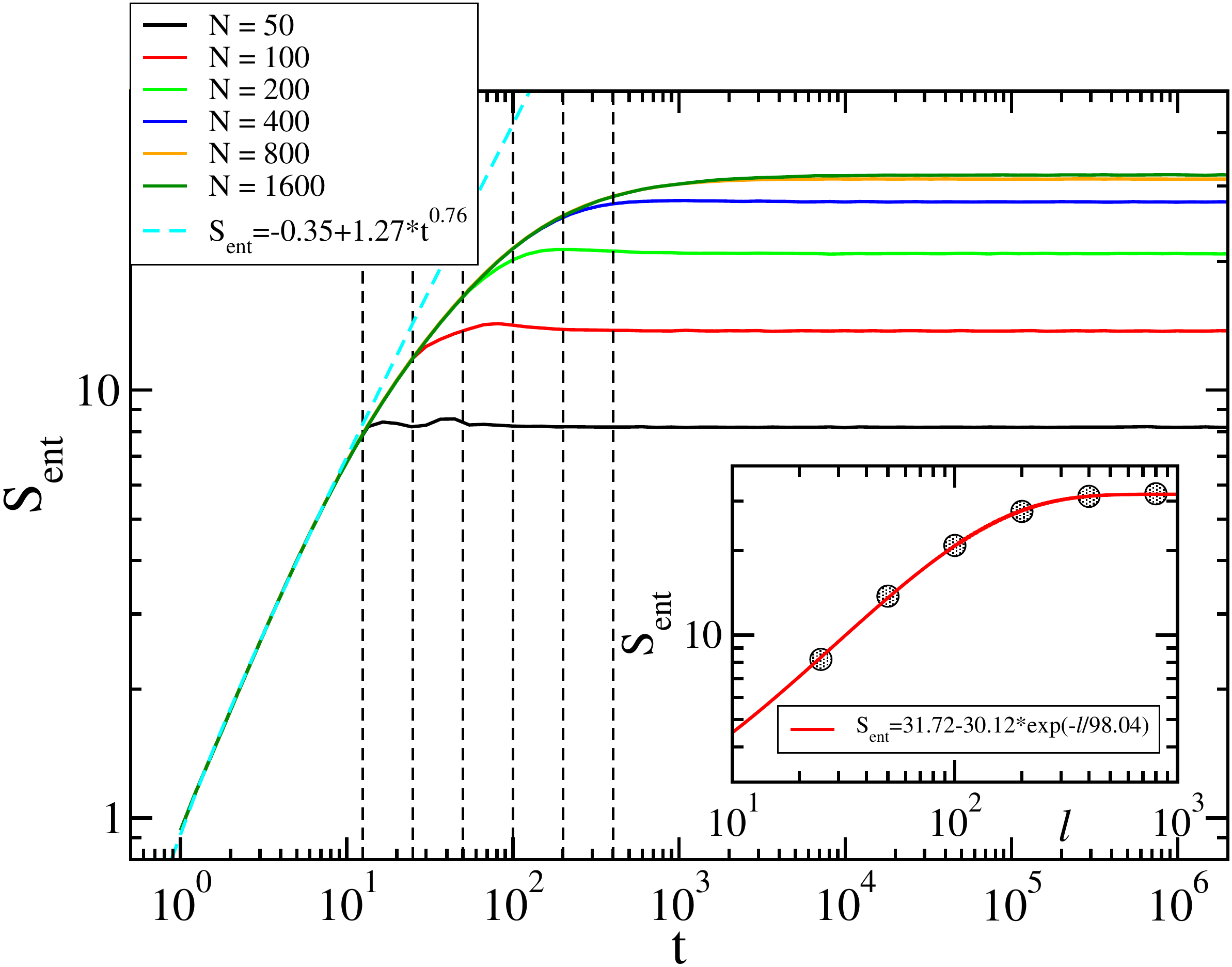}
\caption{(Color online) $S_{\rm ent}$ for the open XX chain with box
  potential disorder $W=1.0$. The dashed vertical lines denote the
  crossover scale $t^*=\ell/2$ in the clean system. Inset: Saturation
  value versus block length $\ell=N/2$.}
\label{Fig3}
\end{figure}

\begin{figure}[ht!]
\includegraphics*[width=0.99\columnwidth]{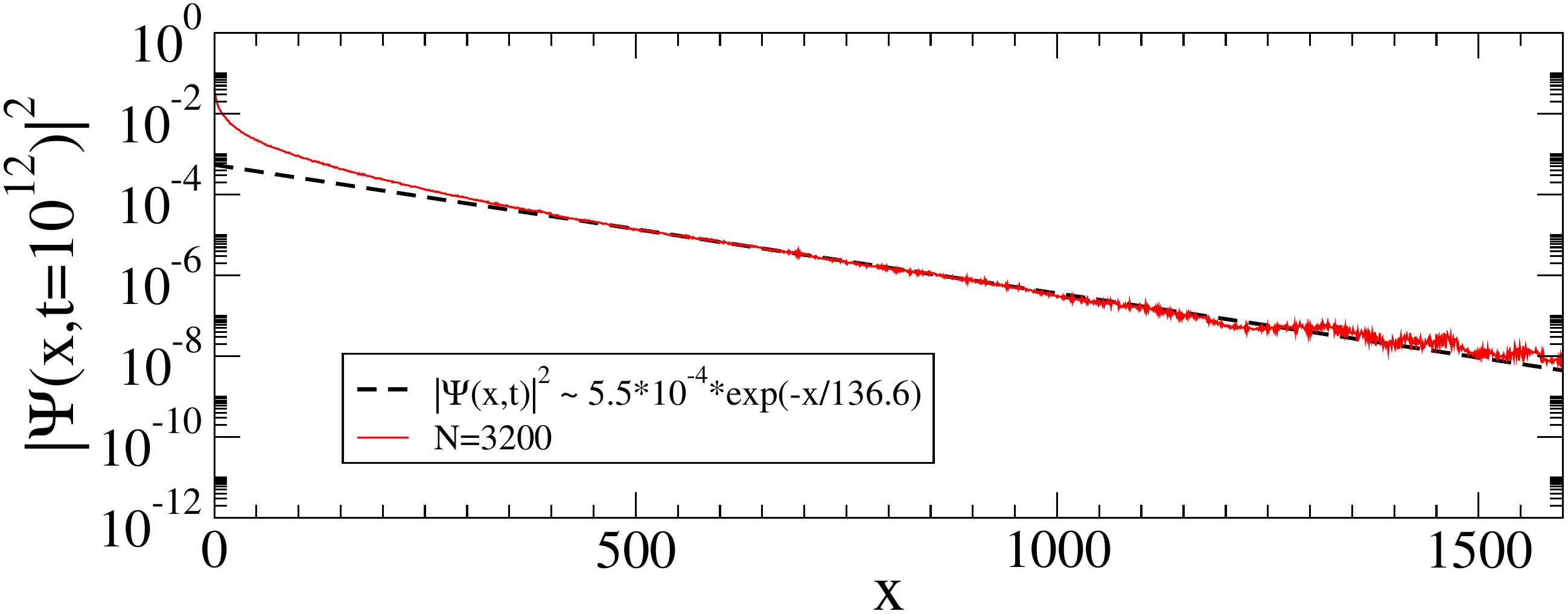}
\caption{(Color online) Disorder averaged stationary probability
  distribution $|\Psi(x,t\to\infty)|^2$ for a single particle
  initially located in the middle of a chain ($x=0$) with $N=3200$
  sites and disorder $W=1.0$.}
\label{Fig4}
\end{figure}

Next, we consider the case $W=1.0$ where we are able to investigate
both regimes, $\xi_{\rm loc}> \ell$ and $\xi_{\rm loc}< \ell$. The
results of exact diagonalizations are shown in Fig.~\ref{Fig3}.
The initial increase of $S_{\rm ent}$ roughly follows a power law with
an exponent smaller than $1$. For $N\lesssim 200$ the crossover scale
$t^*=\ell/2$ is still relevant and marks a deviation from the
entanglement curve in the limit $\ell=N/2\to\infty$.  This means that
for $N\lesssim 200$ a significant contribution from disorder
configurations exists where quasiparticles can still be defined,
propagate almost ballistically, and reach the end of the block at time
$t^*$.  For $N=800$ and $N=1600$ we have reached the regime where
$\xi_{\rm loc}< \ell$ and the results are becoming independent of
block size $\ell$. The inset shows that the saturation values can be
fitted by a simple exponential form. The length in the exponential fit
should be interpreted as being roughly the localization length,
$\xi_{\rm loc}\approx 100$.  This picture is confirmed by a
calculation of the stationary probability distribution
$|\Psi(x,t\to\infty)|^2$ of a single particle which is located at the
center of the chain at time $t=0$.  At long distances the decay is
exponential, $|\Psi(x,t\to\infty)|^2\sim \exp(-x/\xi_{\textrm{loc}})$,
with a fit yielding $\xi_{\textrm{loc}}\sim 136$, see Fig.~\ref{Fig4}.
The saturation value of the entanglement entropy as a function of
block length $\ell$ thus allows to obtain an estimate for the
localization length $\xi_{\rm loc}$.  Here, we expect that the
localization length only weakly depends on energy so that typical and
mean correlation length are of similar magnitude. 

\begin{figure}[ht!]
\includegraphics*[width=0.99\columnwidth]{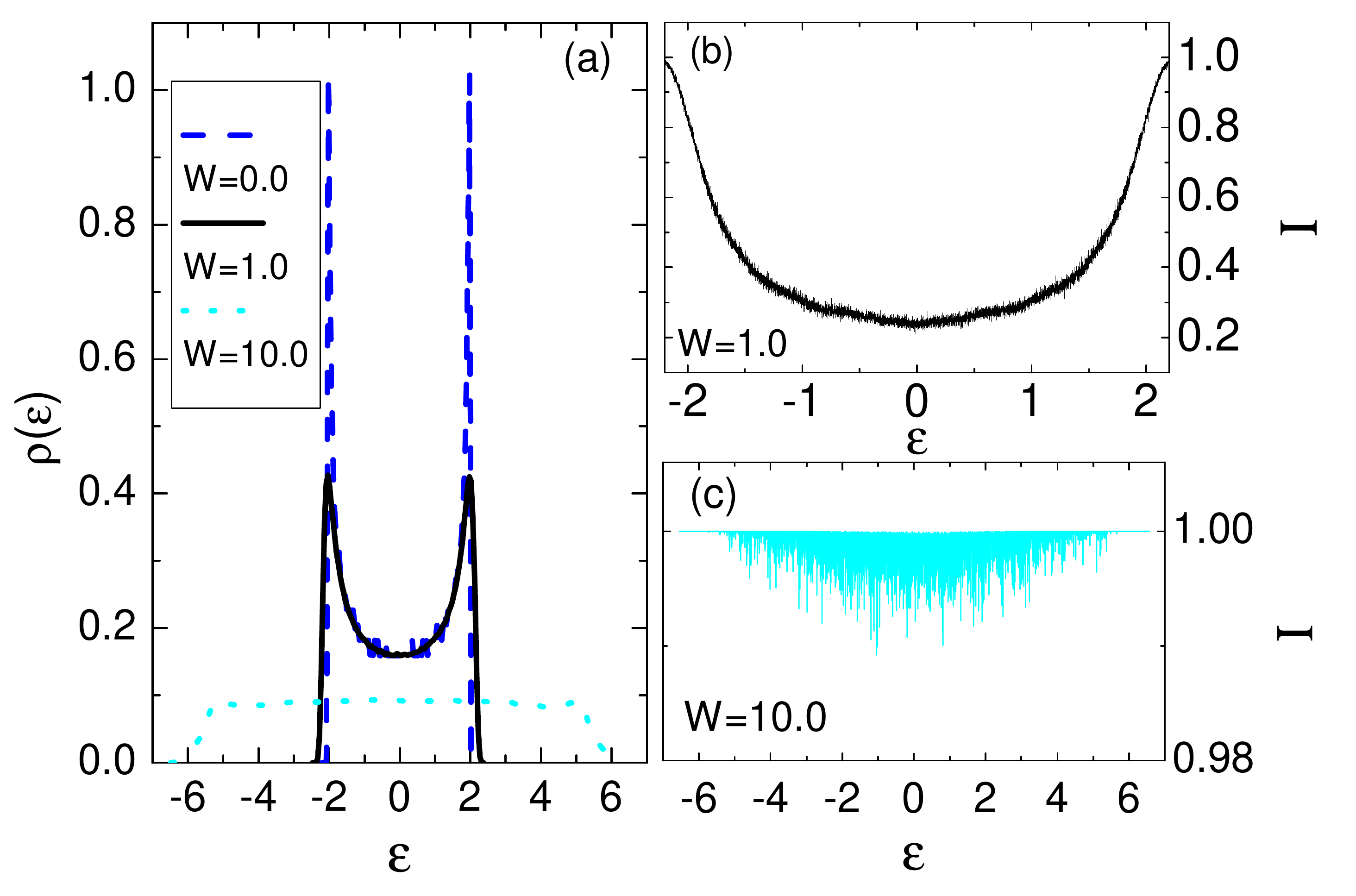}
\caption{(Color online) XX model with potential disorder $W$ and
  $N=1000$ using $2000$ samples: (a) DOS for $W=0.0,\,1.0$, and
  $W=10.0$ using a bin size $\Delta\varepsilon =|\varepsilon_{\rm
    max}|/100$. (b,c) Localization measure $I_\varepsilon$,
  Eq.~\eqref{IPR}, with $m=10$.}
\label{Fig5}
\end{figure}

That the XX model with potential disorder is indeed a generic
disordered system where a delocalization transition as a function of
energy does not occur, is supported further by an analysis of the
spectral properties of the system. We define the density of states
(DOS) as
\begin{equation}
\label{DOS}
\rho(\varepsilon)=\frac{\Delta N}{N\Delta\varepsilon}
\end{equation}
where $\Delta\varepsilon$ is the size of an energy bin, and $\Delta N$
is the number of single-particle states in this bin. The DOS for $W=1$
is shown in Fig.~\ref{Fig5}(a) and is peaked near the edges of the
single-particle spectrum. For $W=10$ the DOS becomes almost box
shaped. To analyze how localized a particular single-particle
eigenstate is we define a localization measure 
\begin{equation}
\label{IPR}
  I_\varepsilon=\sum_{l=-m}^m |\phi_\varepsilon(x_0+l)|^2
\end{equation}
where $\phi_\varepsilon$ is the single-particle eigenfunction with
eigenenergy $\varepsilon$ and $\max_{x\in [1,N]}
|\phi_\varepsilon(x)|^2=|\phi_\varepsilon(x_0)|^2$. I.e.,
$I_\varepsilon$ measures how probable it is that the particle in
eigenstate $\phi_\varepsilon$ is located in an interval
$[x_0-m,x_0+m]$ around the position of the maximum in the probability
distribution. Note that Eq.~\eqref{IPR} is not the inverse
participation ratio. We find that the latter does not always give a
clear picture, in particular, for the critical systems studied later.

If we keep $m$ fixed and send $N\to\infty$ then
$I_\varepsilon\to 0$ if the state is delocalized. For $W=1,10$ and
$m=10$ we find that $I_\varepsilon$ is converged for system size
$N=1000$ and nonzero for all energies, see Fig.~\ref{Fig5}(b,c). This
confirms that all eigenstates are indeed
localized. 

\begin{figure}[ht!]
\includegraphics*[width=0.99\columnwidth]{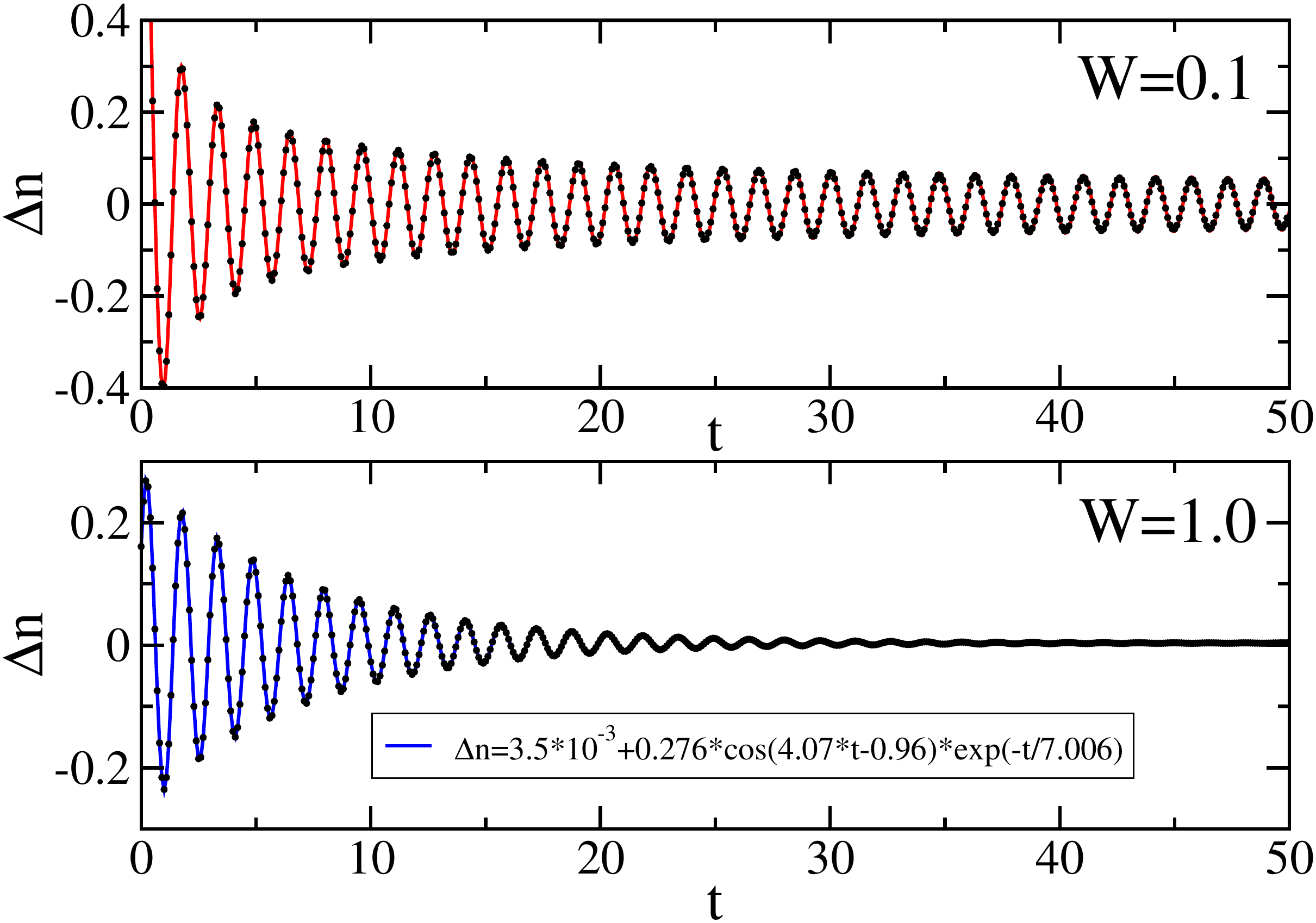}
\caption{(Color online) Density wave order parameter $\Delta n(t)$.
  The circles denote the numerical data. For weak disorder $W=0.1$ the
  result closely resembles the clean case (solid line). For $W=1.0$
  the data are well fitted by an exponential decay to a finite value,
  Eq.~\eqref{fit} (solid line).}
\label{Fig6}
\end{figure}

Of experimental interest is the time evolution of the density wave
order parameter 
\begin{equation}
\label{DWO}
  \Delta n(t)= \frac{2}{N}\sum_j (-1)^j \langle n_j \rangle(t)
\end{equation}
starting from the density wave initial state, Eq.~\eqref{Neel}. For
cold atomic gases $\Delta n(t)$ can be measured
directly.\cite{BlochMBL,PertotSheikhan,BrownWyllie} In the clean case
we obtain, after taking the thermodynamic limit,
\begin{equation}
\Delta n(t)= J_0(4t)\sim (2\pi t)^{-1/2}\cos(4t-\pi/4) 
\end{equation}
where $J_0$ is the Bessel function of the first kind and we have used
the asymptotic expansion at long times. For weak disorder $W=0.1$ and
short times, the order parameter closely follows the time evolution in
the clean case, see Fig.~\ref{Fig6}. The decay in the time interval
shown is therefore well described by a power law $\Delta n\sim
1/\sqrt{t}$ with deviations being expected at longer times.
In contrast, we find an exponential decay 
\begin{equation}
\label{fit}
\Delta n(t) = \Delta n_0 +a\cos(\Omega t-\phi)\exp(-t/\tau) 
\end{equation}
to a finite value $\Delta n_0\approx 4\cdot 10^{-3}$ for disorder
strength $W=1.0$. Thus disorder prevents a full dephasing of $\Delta
n$ and the nonzero value $\Delta n_0$ in the long-time limit is a
signature of localization.
\begin{figure}[ht!]
\includegraphics*[width=0.99\columnwidth]{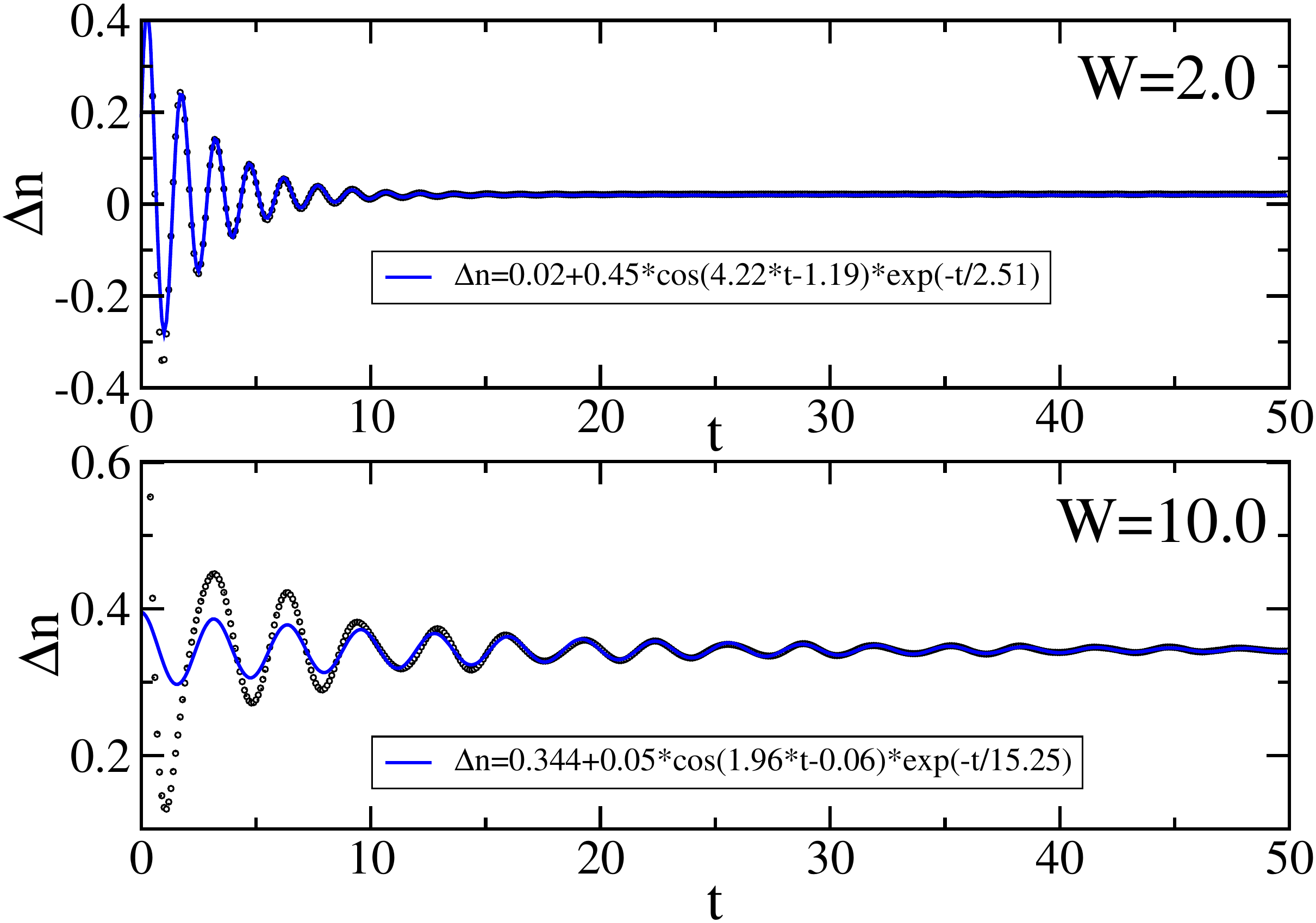}
\caption{(Color online) Density wave order parameter $\Delta n(t)$.
  The decay for $W=2.0$ and $W=10.0$ is still exponential, however,
  the decay rate $\tau$ is a nonmonotonic function of disorder. Lines
  are fits of the asymptotic behavior using Eq.~\eqref{fit}.}
\label{Fig7}
\end{figure}
Note, however, that while the long time mean $\Delta n_0$ is a
monotonic function of disorder $W$, the decay time $\tau$ is not. This
is obvious from the data shown in Fig.~\ref{Fig7} which show very
slowly decaying oscillations for $W=10$. This behavior can be
explained as follows: If $W\gg J$ then, for any disorder
configuration, neighboring sites in the chain exist where the
potential difference $|\mu_i-\mu_{i+1}|$ is much larger than the
hopping $J$. Between these sites the chain is effectively cut and the
finite segments show almost independent dynamics. The case
$W\to\infty$ can be analyzed analytically for a binary disorder
distribution and is discussed in the next subsection.

\subsection{Binary distribution}
Systems with binary potential disorder $\mu_i=\pm W/2$ are ideal for
numerical studies because an exact disorder average in the
thermodynamic limit is possible even in the interacting case. The idea
is to map the system with discrete disorder onto a translationally
invariant system in an enlarged Hilbert space with ancilla sites,
$\mu_i n_i\to \frac{W}{2} n_i\sigma_i$, where $\sigma_i=\pm 1$.
Preparing the ancilla variables $\sigma_i$ in a completely mixed
state, a tracing over the ancillas then gives the exact disorder
average of any local
observable.\cite{ParedesVerstraete,AndraschkoEnssSirker} It is thus
interesting to see if the binary disorder distribution gives
qualitatively the same physics as the box disorder distribution
considered previously.

A disadvantage of mapping the system with binary disorder onto a
transationally invariant system with additional ancilla sites is that
the only entanglement entropy one can then easily calculate is the
entanglement entropy in the enlarged Hilbert space consisting of the
real system {\it and} the ancillas.\cite{AndraschkoEnssSirker} This
entanglement entropy will qualitatively show the same time dependence
as $S_{\rm ent}$ of the system alone because the ancilla sites are
static. A quantitative comparison with the box potential disorder,
however, becomes impossible. In the following, we will therefore not
map the system. Instead, we average over a set of samples as for the
box distribution. We will, however, use the mapping onto a system with
ancillas in section \ref{Interaction} when we discuss interacting
models with binary disorder.

\begin{figure}[ht!]
\includegraphics*[width=0.99\columnwidth]{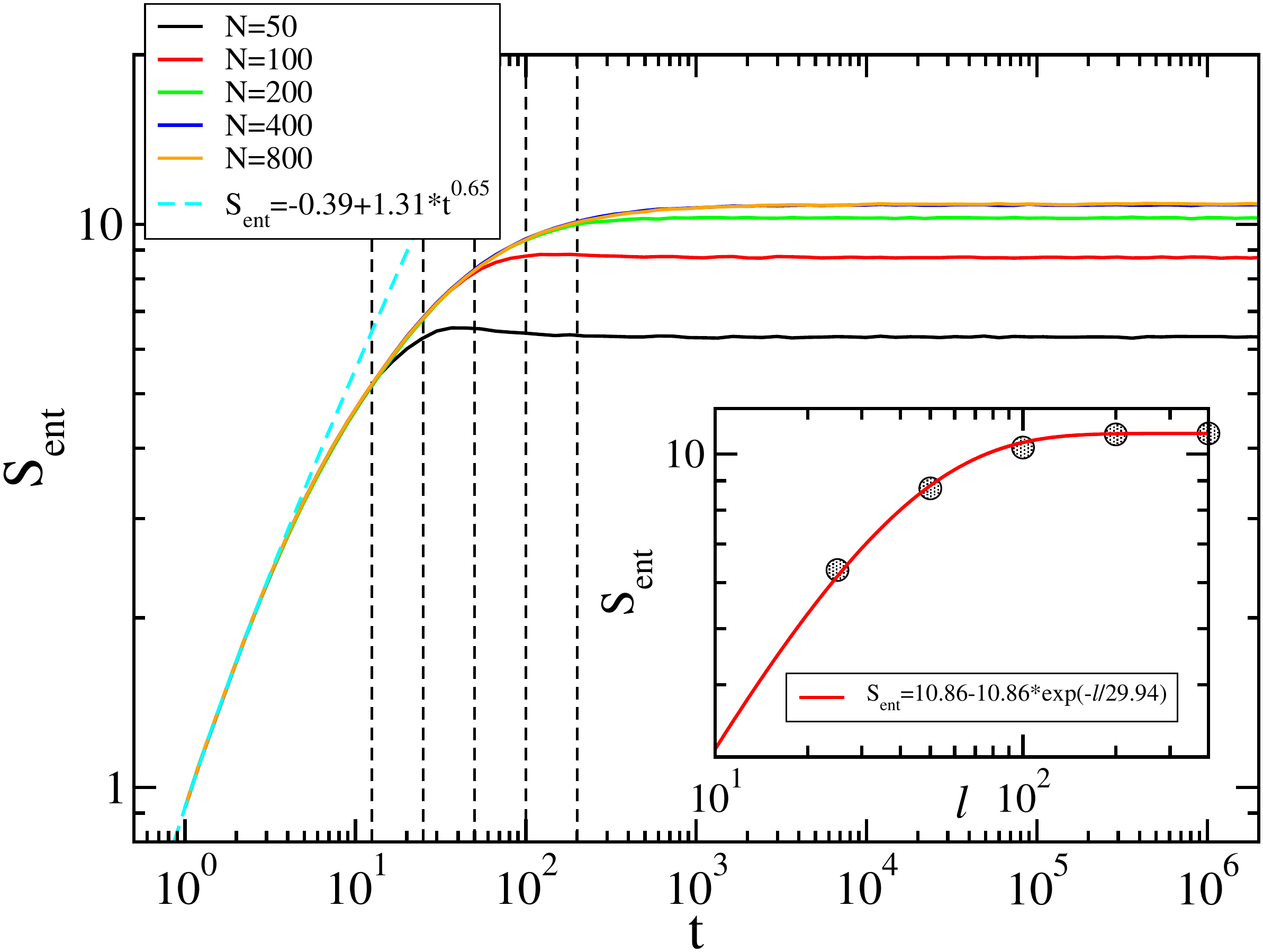}
\caption{(Color online) $S_{\rm ent}$ for the open XX chain with
  binary potential disorder $W=1.0$. The dashed vertical lines denote
  the crossover scale $t^*=\ell/2$ in the clean system. Note that the
  curves for $N=400$ and $N=800$ are almost on top of each other.}
\label{Fig8}
\end{figure}

Results for binary potential disorder $W=1$ are shown in
Fig.~\ref{Fig8} and are qualitatively very similar to the box disorder
case.
We find an increase at small times described approximately by a power
law and a saturation at long times. The approach to saturation in the
limit of infinite block size is again controlled by the localization
length with $\xi_{\rm loc}\approx 30$ in the case considered here.

\begin{figure}[ht!]
\includegraphics*[width=0.99\columnwidth]{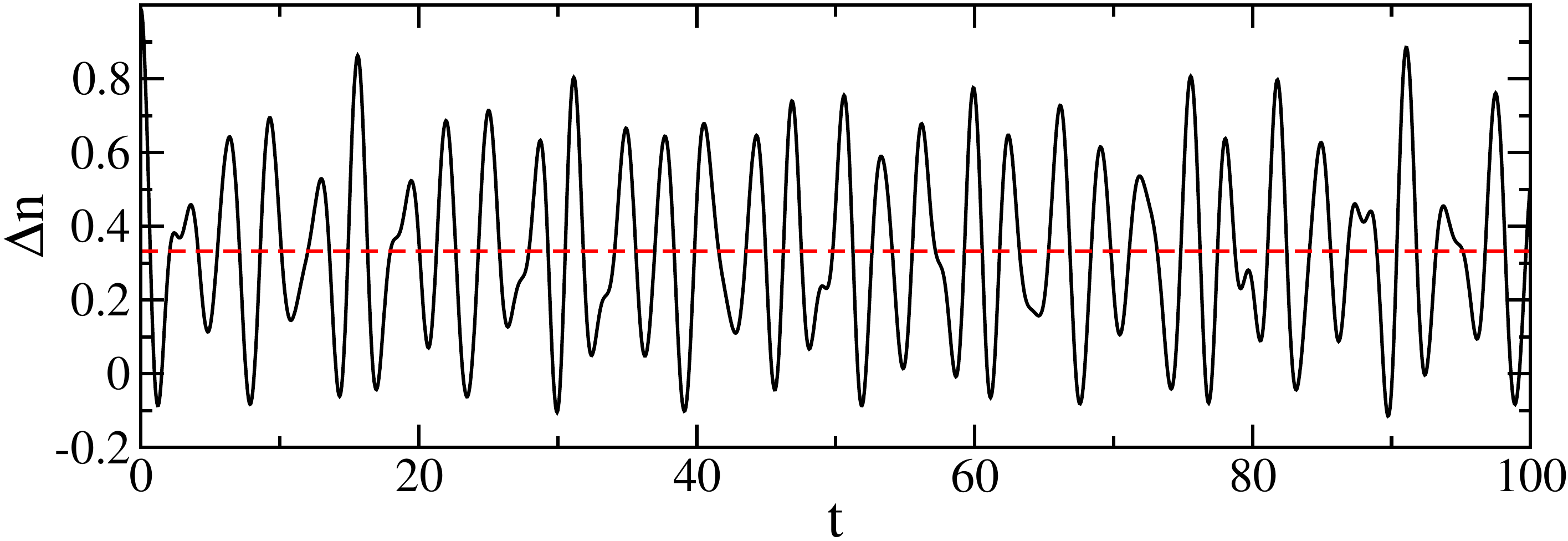}
\caption{(Color online) Density wave order parameter $\Delta n(t)$ for
  infinite binary disorder. The dashed line denotes the long-time mean
  $\overline{\Delta n}=1/3$.}
\label{Fig9}
\end{figure}

Let us finally discuss the case of infinite binary disorder,
$W\to\infty$, following Ref.~\onlinecite{AndraschkoEnssSirker}. In
this case the chain is cut into finite segments of equal potential.
The probability of finding a segment of length $l$ with constant
potential is given by $p(l)=l/2^{l+1}$. For a quench starting from the
initial state \eqref{Neel}, the time evolution of the density wave
order parameter \eqref{DWO} can now be calculated straightforwardly
\begin{equation}
\Delta n(t) = \sum_{l=1}^\infty p_l\Delta n_l(t).
\end{equation}
Here $\Delta n_l(t)$ is the time evolution of a segment of length $l$
with OBC and equal potential on all sites
\begin{equation}
\Delta n_l(t) = \frac{2}{l}\sum_{k=1}^l\exp\left[4it\cos\left(\frac{\pi k}{l+1}\right)\right].
\end{equation}
The oscillations around the mean value $\overline{\Delta
  n}$---determined by the segments with odd length, $\overline{\Delta
  n}=\sum_l p_l \overline{\Delta n}_l
=\sum_{l\;\text{odd}}\frac{2}{l}p_l=\frac{1}{3}$---will therefore
persist for all times, see Fig.~\ref{Fig9}.

\section{XX model with bond disorder}
\label{Bond_disorder}
\begin{figure}[ht!]
\includegraphics*[width=0.99\columnwidth]{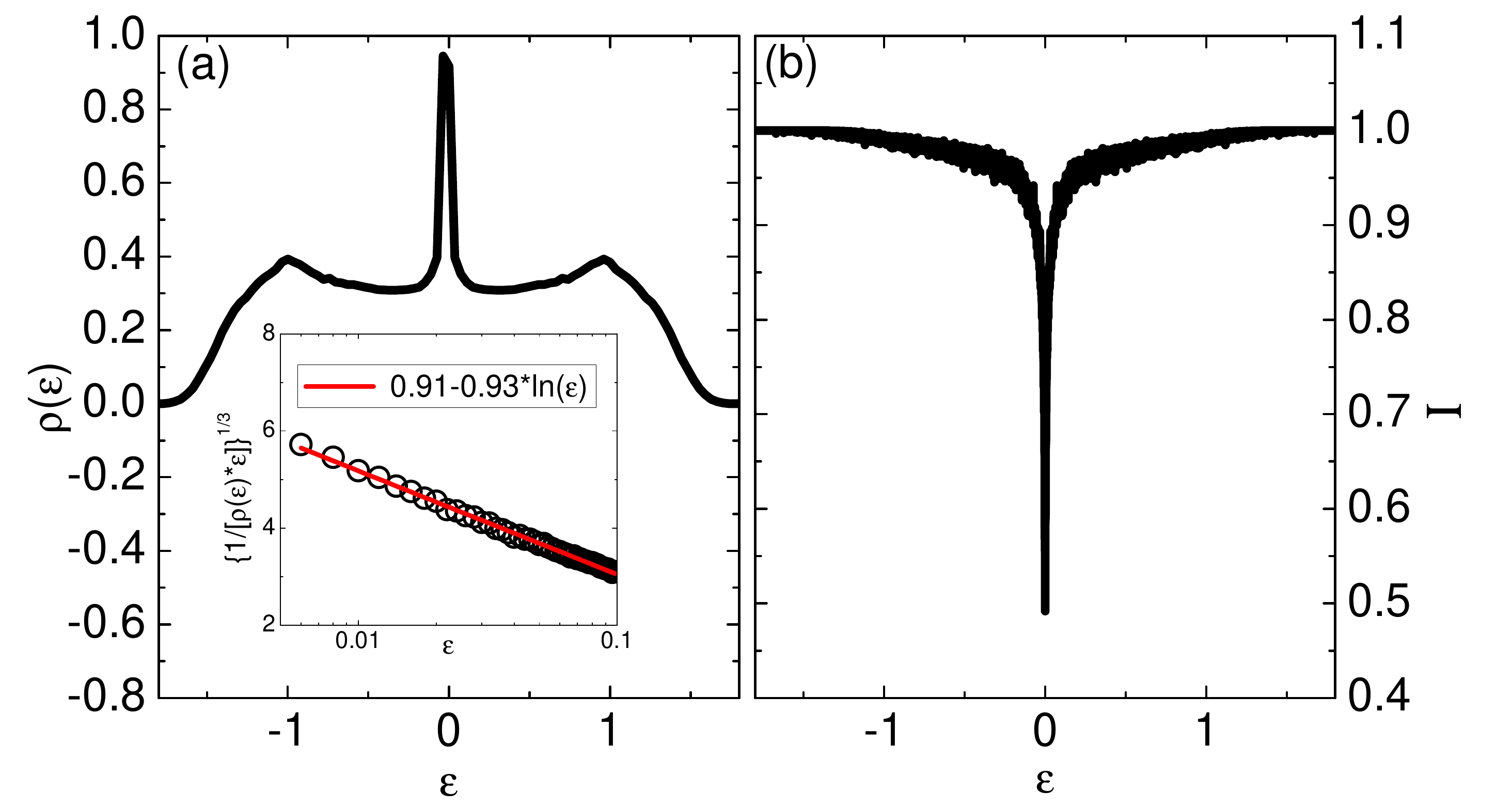}
\caption{(Color online) Open XX chain with $N=1000$ sites and bond
  disorder $J_i\in (0,1)$ using $10000$ samples. (a) DOS based on
  $100$ energy bins. Inset: scaling
  $(|\varepsilon|\rho(\varepsilon))^{-1/3}\sim |\ln\varepsilon^2|$,
  see Eq.~\eqref{DOS_XX}, near $|\varepsilon|\sim 0$ using $1800$ bins.
  (b) $I_\varepsilon$ with $m=10$: delocalization of eigenstates,
  $I_\varepsilon\to 0$ for $\varepsilon \to 0$ in the thermodynamic
  limit. }
\label{Fig10}
\end{figure}

Properties of the XX model, Eq.~\eqref{XX}, with bond disorder $P(J)$,
where $P$ is an arbitrary distribution function of bonds $J_i$, and
$\mu_i\equiv 0$ have been studied first by Eggarter and Riedinger,
Ref.~\onlinecite{EggarterRiedinger}. They found that the density of
states $\rho(\varepsilon)$ shows a divergence at zero energy
\begin{equation}
\label{DOS_XX}
\rho(\varepsilon)\sim 1/|\varepsilon(\ln\varepsilon^2)^3|.
\end{equation}
This behavior is confirmed numerically, see Fig.~\ref{Fig10}(a).  
The divergent density of states at $\varepsilon=0$ is a consequence of
particle-hole symmetry. The localization measure $I_\varepsilon$,
Eq.~\eqref{IPR}, shown in Fig.~\ref{Fig10}(b) furthermore indicates
that the system undergoes a delocalization transition at
$\varepsilon\to 0$. The XX model with bond disorder is therefore an
example for a critical disordered system.

More specifically, it has been shown that the {\it typical}
localization length diverges,\cite{EggarterRiedinger}
\begin{equation}
\label{typ_corr}
\tilde\xi_{\rm loc}\sim |\ln\varepsilon^2|,
\end{equation}
for $\varepsilon\to 0$. Later, it has been emphasized that the spatial
decay of the average Green's function is dominated by the {\it mean}
localization length
\begin{equation}
\label{mean_corr}
\xi_{\rm loc}\sim |\ln\varepsilon^2|^2
\end{equation}
which is much longer than the {\it typical} localization length,
Eq.\eqref{typ_corr}.\cite{Fisher94,BalentsFisher97}

In recent years, a RSRG-X approach has also been used to investigate
the quench dynamics in the XXZ chain with bond disorder leading to the
following three specific predictions for the XX case:\cite{VoskAltman}
(i) The asymptotic growth of entanglement entropy is given by
\begin{equation}
\label{Sent_XX}
S_{\rm ent}(t) \approx \frac{S_p}{3}\ln[\ln(\Omega_0 t)+1/a_0]
\end{equation}
with $S_p=2-1/\ln 2\approx 0.557$. (ii) The entanglement entropy of a
block of length $\ell$ in the thermodynamic limit saturates at long
times
\begin{equation}
\label{Sent_XX2}
S_{\rm ent}(t\to\infty)\approx \frac{S_p}{6}\ln\ell .
\end{equation}
(iii) For a quench starting from the density wave state,
Eq.~\eqref{Neel}, the order parameter \eqref{DWO} decays as
\begin{equation}
\label{Sent_XX3}
\Delta n(t)\approx [a_0\ln(\Omega_0 t) + 1]^{-2}.
\end{equation}
Here $a_0$ and $\Omega_0$ are non-universal constants. In the
following, we will test these scaling predictions using ED data, which
will also allow to analyze the regime of short and intermediate times
where the RSRG-X approach is not applicable.

\subsection{Box distribution}
\begin{figure}[ht!]
\includegraphics*[width=0.99\columnwidth]{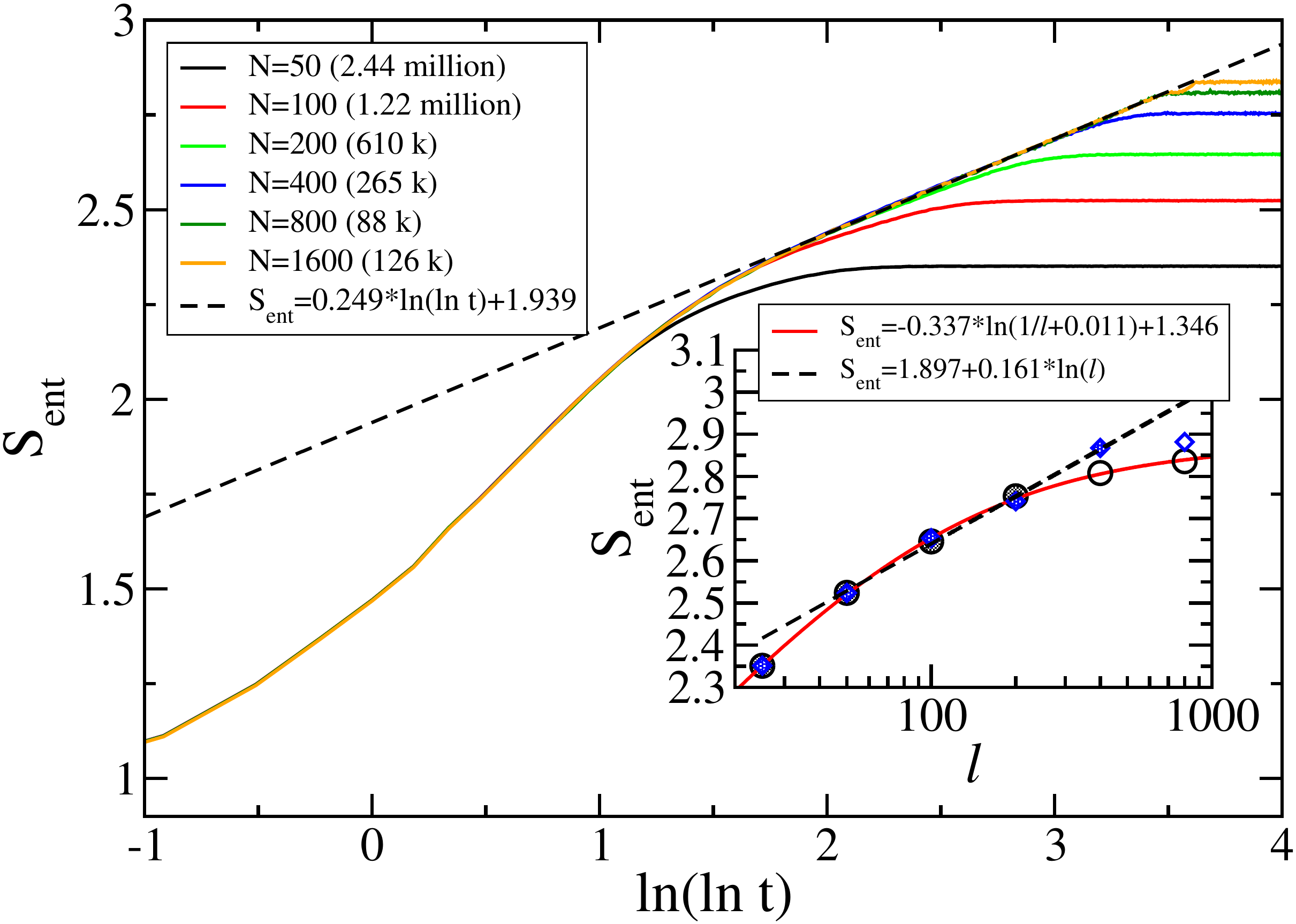}
\caption{(Color online) $S_{\rm ent}(t)$ for the XX model with bond
  disorder $J_i\in (0,1)$ and PBC, obtained using standard double
  precision arithmetics. The numbers in brackets in the legend refer
  to the number of samples. The dashed line is a fit. Inset:
  Saturation value as a function of block length $\ell=N/2$ with
  circles denoting results in double and diamonds results in
  double-double precision.}
\label{Fig11}
\end{figure}

We start by considering chains of length $N$ with PBC and a box
distribution $J_i\in (0,1)$. We keep the block size $\ell=N/2$ fixed.
Data for $S_{\rm ent}(t)$, starting from the initial density wave
ordered state \eqref{Neel}, are shown in Fig.~\ref{Fig11}.
A very well defined $\ln(\ln t)$ scaling is observed over several
orders in time consistent with the RSRG prediction,
Eq.~\eqref{Sent_XX}, however the prefactor deviates from $S_p/3\approx
0.186$. A problem for the numerical calculations are the long times
required to observe saturation. In usual double precision, only $15$
relevant digits are kept. The spectrum of the XX model with bond
disorder contains, however, many very small eigenvalues
$\varepsilon_i$ (see Fig.~\ref{Fig10}). These eigenvalues and the
corresponding eigenstates become relevant for the time evolution if
$t\varepsilon_i \gtrsim 1$. Since eigenvalues smaller than $10^{-14}$
and their corresponding eigenvectors in a spectrum which also contains
eigenvalues of order $1$ cannot be determined accurately in double
precision, we expect the numerical data in Fig.~\ref{Fig11} to become
unreliable for $\ln(\ln t)\gtrsim 3.5$. An indication of these
numerical problems is a small dip in the curve for $N=1600$ close to
the times where the entanglement entropy starts to saturate.

The saturation value $S_{\rm ent}(\ell=N/2,t\to\infty)$ is shown in
the inset of Fig.~\ref{Fig11} on a logarithmic scale. A naive fit of
the double precision data seems to point to a saturation,
$\lim_{\ell\to\infty}S_{\rm ent}(\ell=N/2,t\to\infty)\approx 2.9$.
However, there are two problems with such a fit: (i) For $N=50$ the
regime where $S_{\rm ent}(t)\sim\ln\ln t$ is never reached. The
related scaling of the saturation value $S_{\rm
  ent}(\ell,t\to\infty)\sim\ln\ell$ therefore cannot be expected to
hold. System sizes $N\gtrsim 100$ are required. (ii) Only systems with
$N\leq 400$ show a saturation at times $t\leq 10^{14}$. The saturation
values for $N=800,\, 1600$ obtained in double precision are not
reliable. In order to overcome problem (ii) we have therefore also
performed calculations in double-double precision where 30 relevant
digits are kept and results for $t\lesssim 10^{30}$, corresponding to
$\ln\ln t \lesssim 4.2$, are expected to be reliable. Here we have
used the linear algebra package MPACK which makes calculations in
fixed double-double or quadruple precision as well as in arbitrary
precision possible.\cite{MPACK} Such calculations lead to a
significantly larger saturation value for $N=800$ (using $10000$
samples for each point in time and averaging over several times of
order $\ln\ln t\sim 4$). However, while the saturation value for
$N=1600$ also increases we still find eigenvalues in many samples
which are effectively zero even in double-double precision. Here
quadruple precision would be required to obtain the saturation value
reliably.  Unfortunately, the computational costs for such
calculations are becoming prohibitive.  The data for the saturation
values which are reliable in double-double precision can be fitted by
a logarithm, see inset of Fig.~\ref{Fig11}.  However, given the
limited range of accessible system sizes this fit is not fully
convincing. We also note that the ratio of the prefactor of the
$\ln\ln t$ scaling and the possible $\ln\ell$ scaling is $a/b=1.5$
instead of $2$ as expected from Eqs.~\eqref{Sent_XX} and
\eqref{Sent_XX2}.

\begin{figure}[ht!]
\includegraphics*[width=0.99\columnwidth]{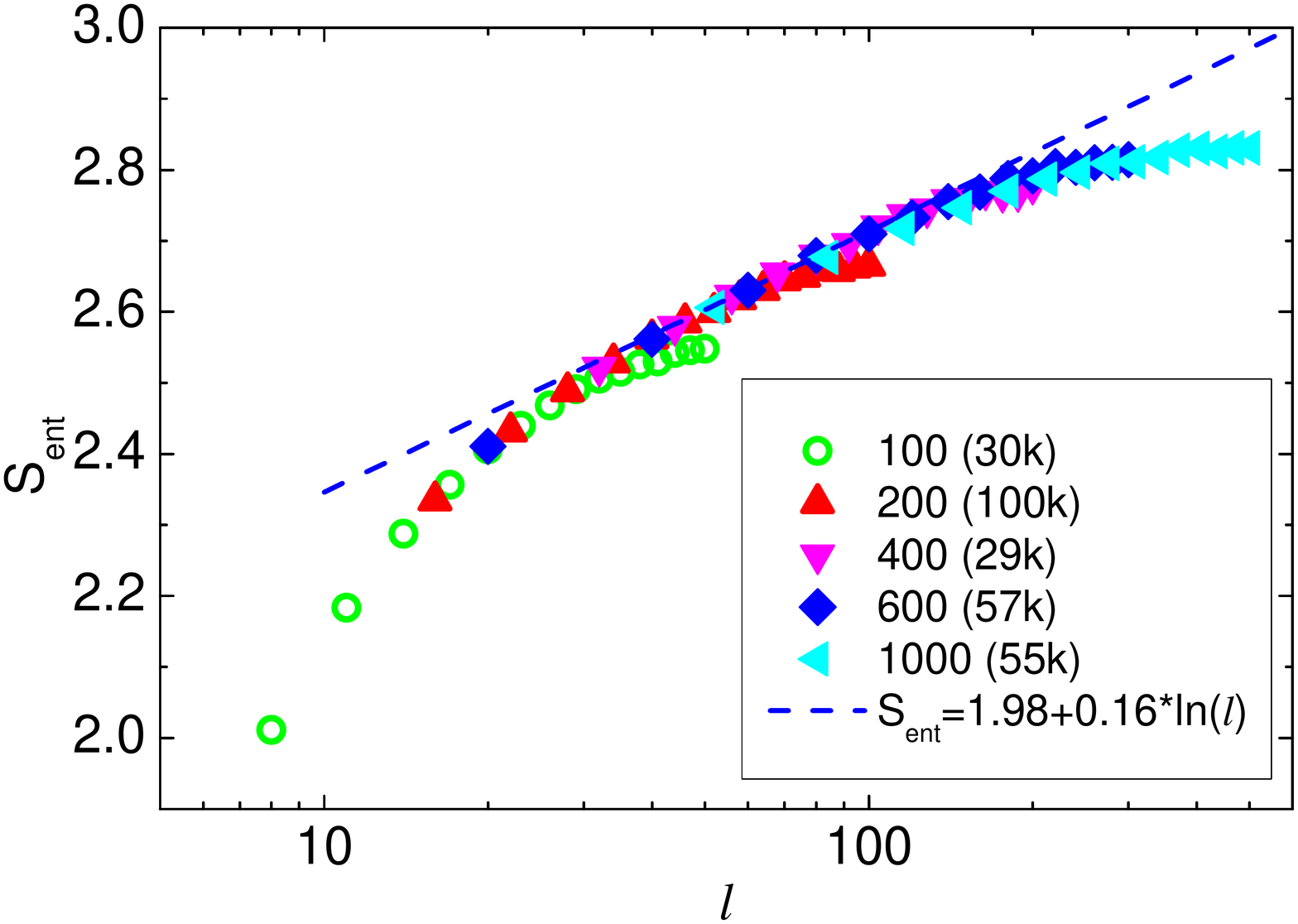}
\caption{(Color online) Double precision results: Saturation value
  $S_{\rm ent}(\ell,t\to\infty)$ for a quench from the initial state
  \eqref{Neel} in the periodic XX chain with $J_i\in (0,1)$ and
  various chain lengths $N$ and block sizes $\ell$ as indicated. The
  dashed line is a fit representing potential logarithmic scaling.}
\label{Fig12}
\end{figure}

Another possible issue is that Eq.~\eqref{Sent_XX2} applies to a block
in an infinite system while the block in our calculations is exactly
half of the system. In Fig.~\ref{Fig12} we therefore present, in
addition, results where we investigate the scaling of the saturation
value more systematically as a function of block size $\ell$.
Again, we expect that small blocks are not in the scaling regime and
that the data for $N>400$ become unreliable due to the limited
numerical precision. Nevertheless, deviations from a logarithmic
scaling (indicated by the dashed line in Fig.~\ref{Fig12}) appear to
be also present for block sizes $\ell\sim 100 - 200$ and $N\leq 400$
where we expect to be in the scaling regime and the numerical data to
be reliable. In conclusion, our numerical data are insufficient to
show unambiguously whether or not the logarithmic scaling,
Eq.~\eqref{Sent_XX2}, holds. Note, that a saturation
$\lim_{\ell\to\infty}\lim_{t\to\infty}S_{\rm
  ent}(\ell,t)=\mbox{const}$ would also imply that $S_{\rm
  ent}\sim\ln\ln t$ is just a transient and not the true scaling
behavior.

\begin{figure}[ht!]
 \includegraphics*[width=0.99\columnwidth]{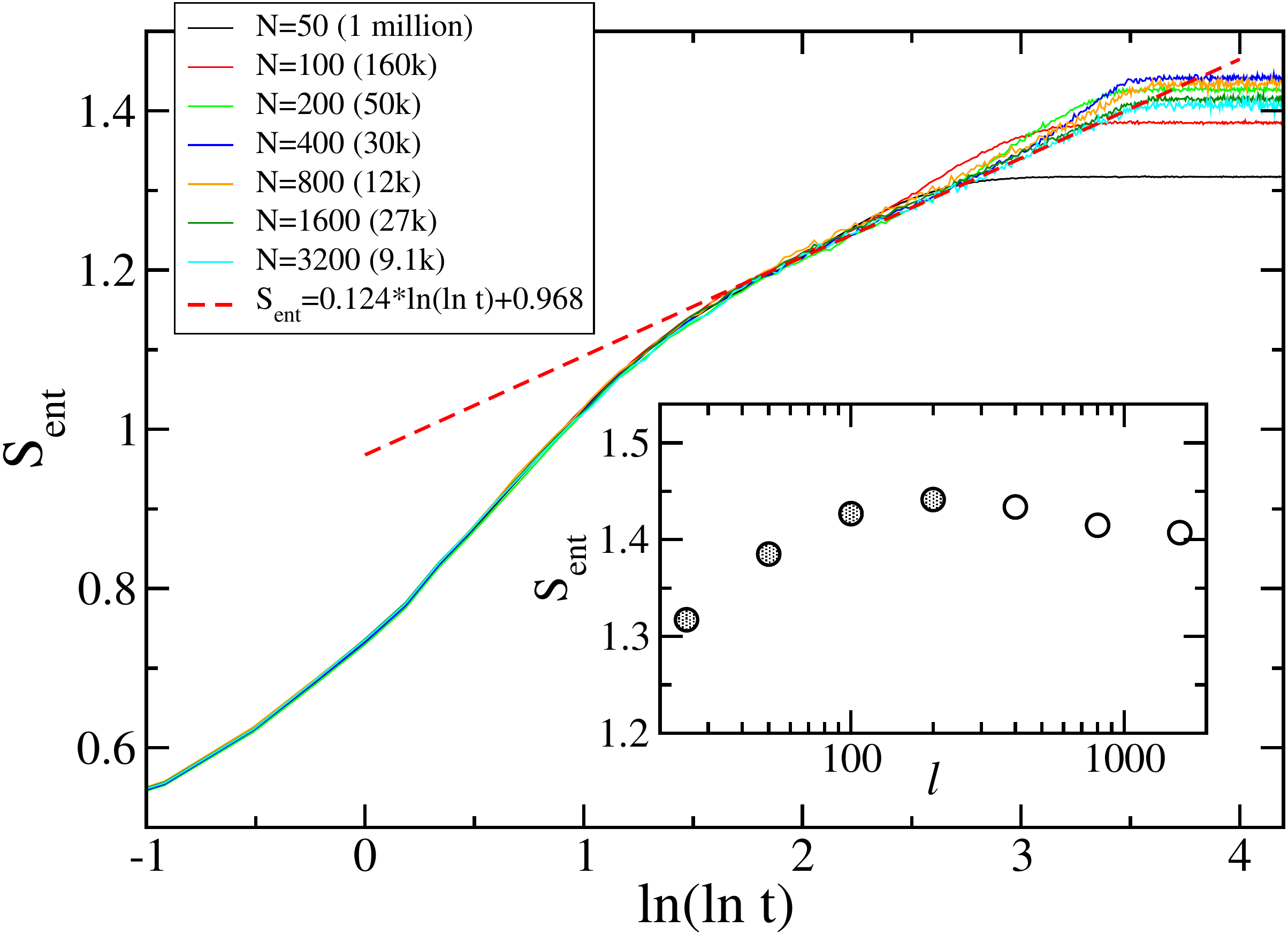}
 \caption{(Color online) Same as Fig.~\ref{Fig11} but with open
   instead of periodic boundary conditions. The $N=3200$ data are
   fitted and a $\ln(\ln t)$ scaling is observed. Inset: Saturation
   value as a function of block length $\ell=N/2$.}
 \label{Fig13}
 \end{figure}

We find an even more intricate behaviour analysing the numerical data
for the same set of parameters as in Fig.~\ref{Fig11} but with open
instead of periodic boundary conditions. The results of calculations
in double precision are shown in Fig.~\ref{Fig13}.
 The entanglement entropy $S_{\rm ent}(t)$ at late times is now a
 non-monotonic function of block size $\ell=N/2$, in particular, the
 saturation value $S_{\rm ent}(\ell)$ appears to have a maximum at
 $\ell=N/2\approx 200$ and to approach a finite value for
 $\ell\to\infty$ from above, see inset of Fig.~\ref{Fig13}. Note,
 however, that the data for $\ell=N/2>200$ at long times are again
 affected by problems with the numerical precision. Nevertheless, the
 data for $N=50-200$ are all in the scaling regime and appear
 reliable, yet a clear logarithmic scaling is not observed. The
 non-monotonic behavior in time is apparently a consequence of the
 interplay of finite block size and boundary contributions. In the
 regime where $S_{\rm ent}\sim\ln(\ln t)$ we find to a very good
 accuracy that $S_{\rm ent}^{\rm PBC}(t)=2S_{\rm ent}^{\rm OBC}(t)$,
 see the fits in Fig.~\ref{Fig11} and Fig.~\ref{Fig13}. For large
 system sizes the two cuts between the system block of length $\ell$
 and the environment of length $N-\ell$ in the case of periodic
 boundary conditions therefore each give a contribution equivalent to
 the single cut for open boundaries.

\begin{figure}[ht!]
\includegraphics*[width=0.99\columnwidth]{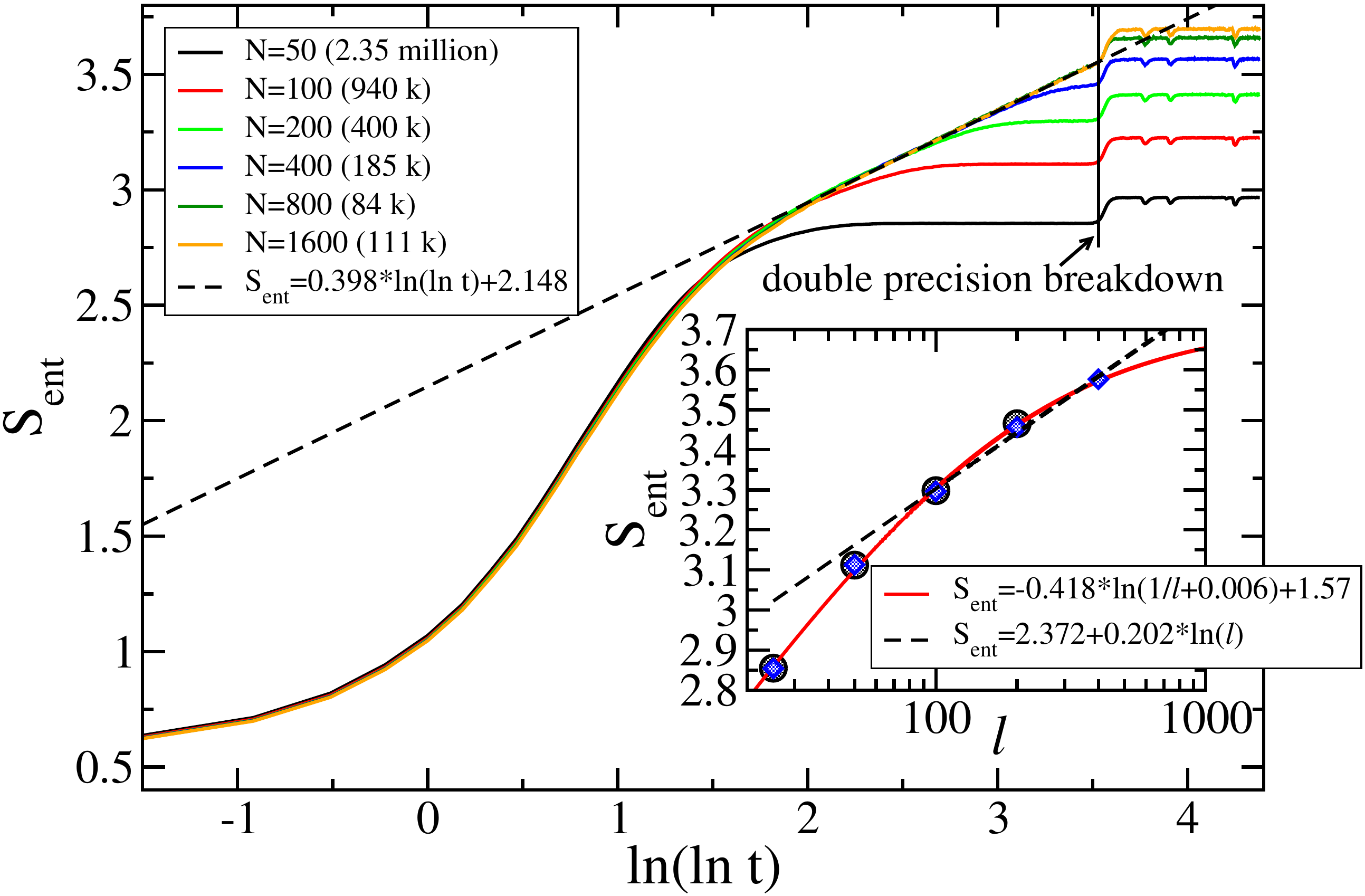}
\caption{(Color online) Same as Fig.~\ref{Fig11} but quenching from a
  random initial product state instead of the density-wave state
  \eqref{Neel}. Inset: Saturation value with circles denoting results
  in double and diamonds results in double-double precision.}
\label{Fig11b}
\end{figure}

 The RSRG-X predictions, Eqs.~\eqref{Sent_XX}-\eqref{Sent_XX3}, and
 our numerical analysis so far are for a quench starting from the
 density-wave state \eqref{Neel}. This is a special state and the
 observed growth of the entanglement entropy might therefore not be
 generic.\cite{VoskAltman,VasseurFriedman} To investigate this point
 we show in Fig.~\ref{Fig11b} additional data sets where we quench
 from a random initial product state.
We do again find a very clear $\ln(\ln t)$ scaling in a certain time
window, however, the prefactor is quite different from the one found
for the quench from the density wave state. At $\ln(\ln t)\approx 3.5$
the double precision calculations break down leading to an artificial
jump in $S_{\rm ent}(t)$. We can therefore only extract the saturation
values for $\ell=N/2\leq 200$ in double precision and for
$\ell=N/2\leq 400$ in double-double precision, see the inset of
Fig.~\ref{Fig11b}. The saturation values could be consistent either
with a $\ln\ell$ scaling or with a saturation $S_{\rm ent}\approx 3.7$
for $\ell\to\infty$. Assuming that a logarithmic scaling does hold,
the fits in Fig.~\ref{Fig11b} yield a ratio $a/b=1.97$ which is close
to $2$ as expected based on the RSRG analysis.

\begin{figure}[ht!]
\includegraphics*[width=0.99\columnwidth]{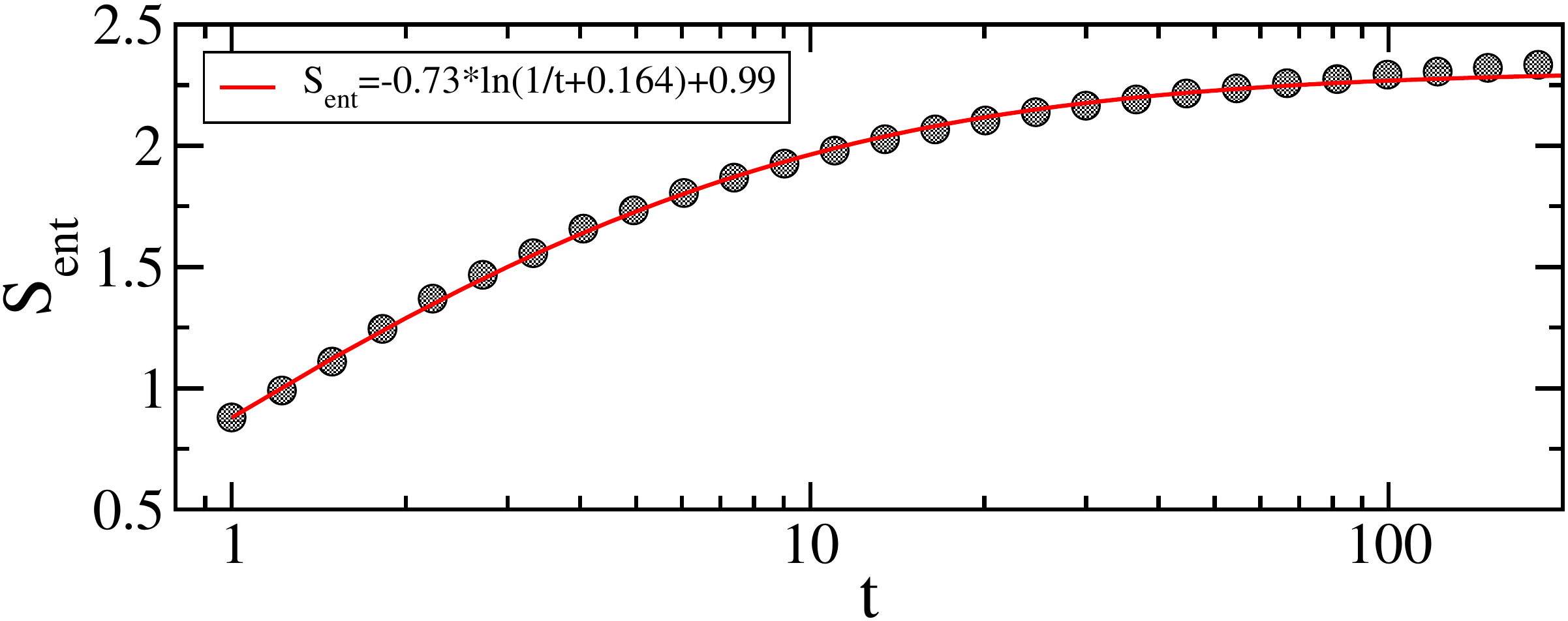}
\caption{(Color online) $S_{\rm ent}(t)$ for the XX model with bond
  disorder $J_i\in (0,1)$, PBC, and quenching from the density wave
  state. The line is a fit, showing that the initial increase is
  almost logarithmic in time.}
\label{Fig14}
\end{figure}

Finally, we analyze the short time behavior of $S_{\rm ent}(t)$ for
periodic boundary conditions quenching from the density wave state,
see Fig.~\ref{Fig14}.
We find that the initial increase can be fitted by the function
displayed in Fig.~\ref{Fig14} and appears almost logarithmic when
viewed in a limited time interval. This should be kept in mind when
analyzing numerical data for small interacting systems where a
logarithmic increase of the entanglement entropy is usually understood
as a signature of a many-body localized phase.\cite{BardarsonPollmann}
The data in Fig.~\ref{Fig14} show that a very similar behavior in a
restricted time window can also be observed in a critical disordered
noninteracting system. This is consistent with previous findings by
Chiara {\it et al.}, Ref.~\onlinecite{ChiaraMontangero06}.

As for the potential disorder case, we also consider the decay of the
density wave order parameter \eqref{DWO} in a quench from the initial
state \eqref{Neel} for bond disorder.
\begin{figure}[ht!]
\includegraphics*[width=0.99\columnwidth]{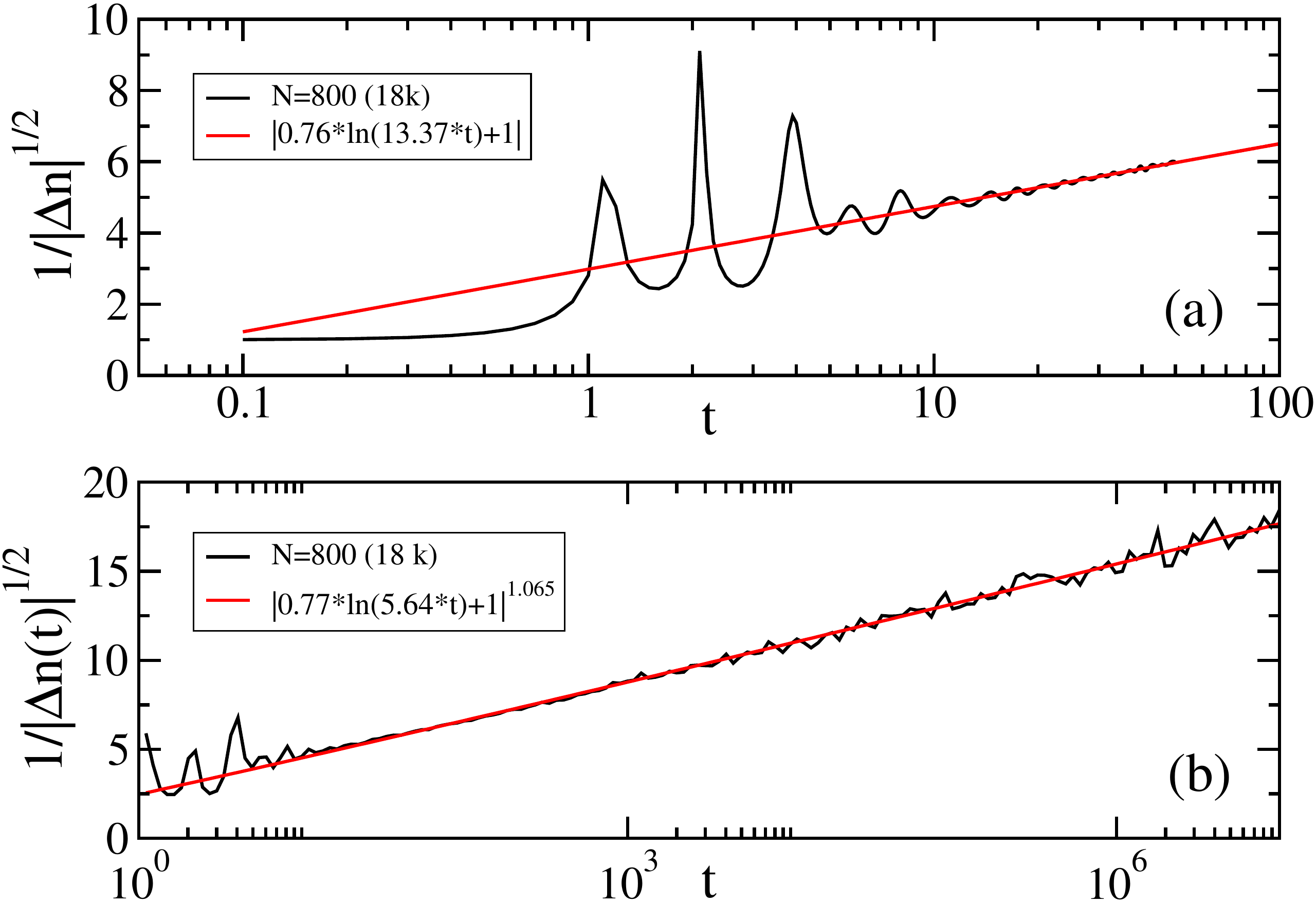}
\caption{(Color online) Decay of $\Delta n(t)$, Eq.~\eqref{DWO}, in
  the XX model with bond disorder $J_i\in [0,1)$, $N=800$ sites, PBC,
  and averaging over $18000$ samples. (a) Short time behavior, and (b)
  long time decay.}
\label{Fig15}
\end{figure}
In Fig.~\ref{Fig15} we plot $1/\sqrt{|\Delta n|}$ as a function of
time. The fit in Fig.~\ref{Fig15}(a) for times $t\in [20,50]$ shows
that the decay is consistent with the RSRG-X prediction,
Eq.~\eqref{Sent_XX3}. Here we have kept the exponent fixed. In
Fig.~\ref{Fig15}(b) a similar fit with $t\in [20,10^6]$ is presented
where also the exponent is used as a fitting parameter. The result is
consistent with the fit at short times. Note that Eq.~\eqref{Sent_XX3}
is, strictly speaking, only valid in an infinite system. For any
finite system and any given disorder configuration we will ultimately
expect recurrences. Due to the disorder average and the broad
distribution of couplings $J_i$, however, the asymptotic decay of the
order parameter can be observed over several orders of magnitude in
time in a system with $N=800$ sites only.


\subsection{Binary distribution}
Instead of bond disorder drawn from a box distribution as investigated
in the previous subsection, we will consider here binary bond disorder
$J_i=1\pm \delta$. This case is not only interesting because it allows
for a numerical investigation of infinite chains even if interactions
are added (see Sec.~\ref{Interaction}) but also because the
applicability of the RSRG-X approach seems questionable in this case.
In the RSRG-X, fast dynamics caused by the strongest bonds in the
random chain is eliminated and the couplings between the remaining
degrees of freedom are renormalized. The control parameter is the
ratio of the coupling constants to the two adjacent sites over the
coupling of the strong bond. For binary disorder---where the chain
consists of segments with length $\ell=2,3,\cdots$ with the {\it same}
coupling $J=1+\delta$ or $J=1-\delta$---this ratio will often be equal
to $1$.  Instead, one might therefore start by eliminating segments of
strong bonds beginning with the shortest ones which are responsible
for the fastest dynamics. This will lead to a broader distribution of
couplings for which a standard RSRG approach might again be
applicable. However, long segments with strong couplings $J=1+\delta$
will show an internal slow dynamics not captured in such an approach
so that it is a priori unclear if such a modified RSRG-X description
can become correct asymptotically.

\begin{figure}[ht!]
\includegraphics*[width=0.99\columnwidth]{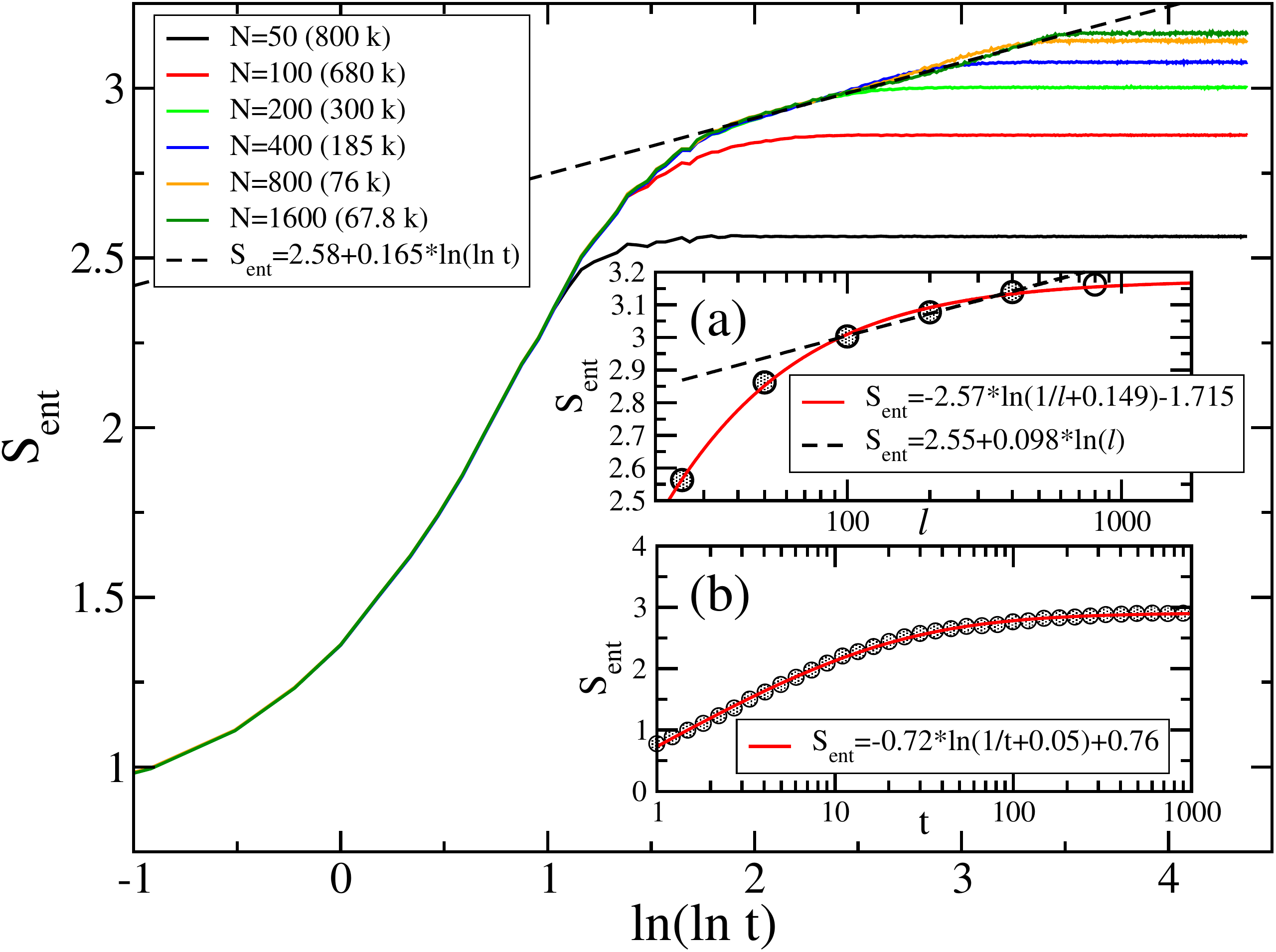}
\caption{(Color online) Main: $S_{\rm ent}(t)$ for the XX model with
  binary bond disorder $\delta=0.4$ and OBC. Fit of the $N=1600$ data
  (dashed line). (a) Saturation value $S_{\rm ent}(\ell,t\to\infty)$,
  and (b) approximately logarithmic scaling of $S_{\rm ent}(t)$ for
  $t\leq 1000$.}
\label{Fig17}
\end{figure}

In Fig.~\ref{Fig17} data for a quench with the initial state
\eqref{Neel}, OBC, and $\delta=0.4$ are shown.
Qualitatively, the results are surprisingly similar to the case of a
box distribution. The initial increase of $S_{\rm ent}(t)$ is again
approximately logarithmic up to times $t\approx 10^3$, see
Fig.~\ref{Fig17}(b). For $t\in [10^4,10^{14}]$ and $N=1600$ we find a
scaling regime with $S_{\rm ent}(t)\sim 0.165\ln(\ln t) + 2.58$. While
this is qualitatively the same behavior as for the box distribution,
the prefactor is different when compared to the fit in
Fig.~\ref{Fig13} ($0.165$ as compared to $0.124$). The saturation
values are reached quicker than for the box distribution and data up
to $\ell=N/2=400$ should be reliable in double precision. All the data
can be fitted by the function shown as solid line in
Fig.~\ref{Fig17}(a) leading to a finite saturation value,
$\lim_{\ell\to\infty}S_{\rm ent}(\ell,t\to\infty)\approx 3.18$, in the
thermodynamic limit. However, the data for $N=50,\,100$ are not in the
scaling regime. Excluding these data a logarithmic fit is also
possible, see the dashed line in Fig.~\ref{Fig17}(a). Again, we cannot
unambiguously show numerically whether or not Eq.~\eqref{Sent_XX2}
holds.

\begin{figure}[ht!]
\includegraphics*[width=0.99\columnwidth]{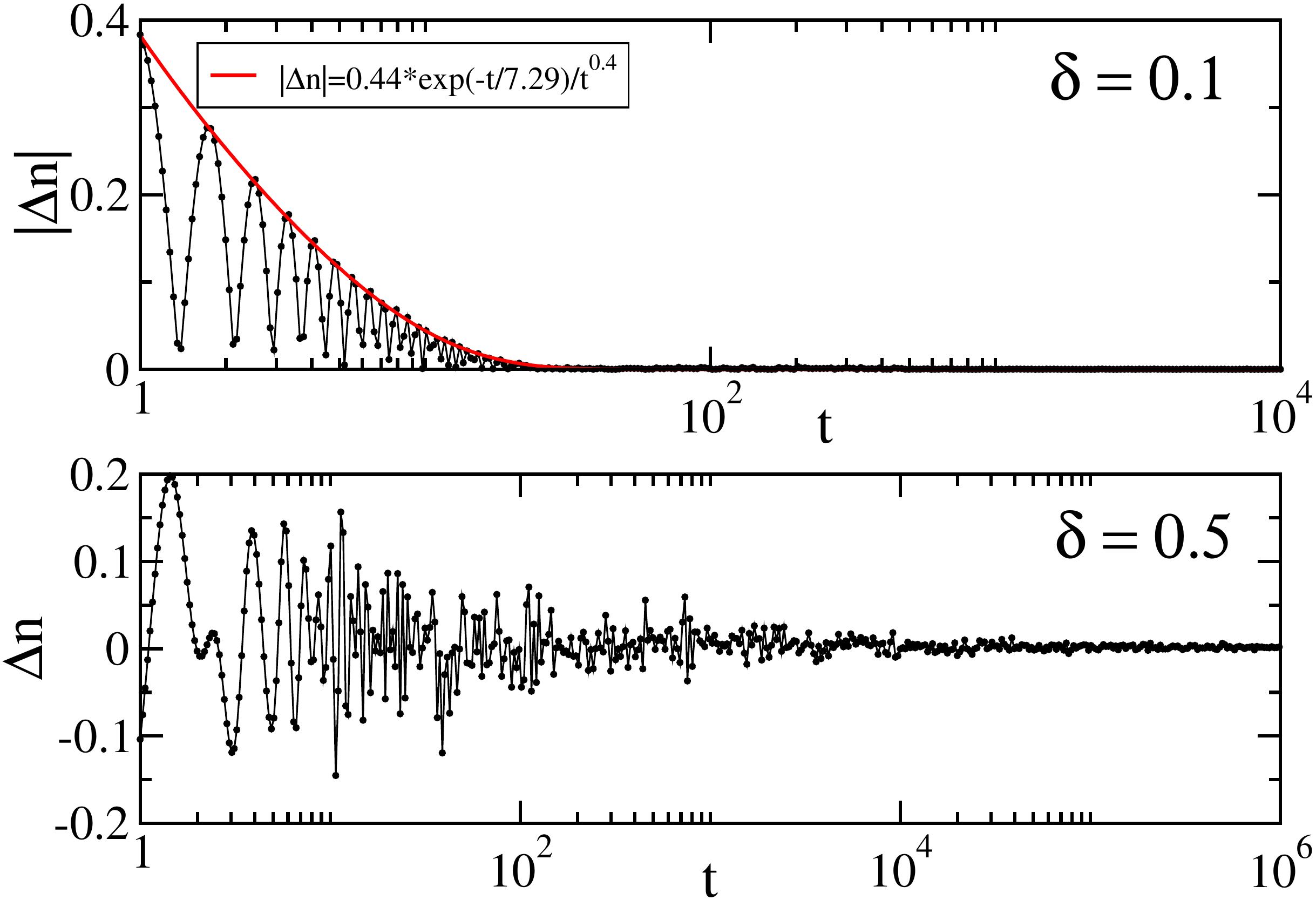}
\caption{(Color online) $\Delta n(t)$ for the XX model with $N=800$
  sites, $40000$ samples, and binary bond disorder $\delta$ as
  indicated. The lines are a guide to the eye.}
\label{Fig_bin_mstagg}
\end{figure}

Also interesting is the time evolution of the density wave order
parameter, Eq.~\eqref{DWO}. For $\delta=1$ the chain separates into
finite segments and $\Delta n(t)$ oscillates around $\overline \Delta
n=1/3$ exactly as for the case of infinite binary potential disorder,
see Fig.~\ref{Fig9}. The only difference is that time is now rescaled
by a factor $2$ because the hopping along the chain segments is given
by $J=1+\delta =2$ instead of $J=1$ for the case of infinite binary
potential disorder. For $0<\delta<1$ we expect $\Delta n(t)$ to decay
for long times in an infinite chain. Note, however, that the weak
bonds introduce a dephasing time scale $t_p=1/(1-\delta)$ which
diverges for $\delta\to 1$. 
In our numerical analysis we therefore concentrate on small and
intermediate strengths of binary disorder. Results for $\Delta n(t)$
with $\delta=0.1$ and $\delta=0.5$ are shown in
Fig.~\ref{Fig_bin_mstagg}.
For $\delta=0.1$ the density wave order decays quickly and the
envelope can be very well fitted by an exponential with an algebraic
correction. For $\delta=0.5$ the density wave order parameter also
decays, however, the strong oscillations make it difficult to observe
any clear scaling. In both cases, the asymptotics \eqref{Sent_XX3}
predicted by RSRG-X is not observed. Note, however, that it is very
difficult to analyze the regime of long times where $|\Delta n|$ is
small and the relative fluctuations are large so that it is impossible
to draw any definite conclusions from the numerical data.

\section{Critical random transverse Ising model}
\label{Ising}
As a second example for a critical random model, we consider the
transverse Ising model
\begin{equation}
\label{Ising_model}
H=-\sum_{i=1}^N \left\{J_i\sigma^x_i\sigma_{i+1}^x +h_i\sigma^z_i\right\} 
\end{equation}
with periodic boundary conditions. This model can be mapped onto a
chain of free fermions of length
$2N$.\cite{Pfeuty,LiebSchultzMattis,IgloiSzatmari,IgloiSzatmari2} The
entanglement entropy can therefore be calculated using the same single
particle algorithms as used previously for the XX model. If $J_i$ and
$h_i$ are drawn independently from the same distribution function then
the model \eqref{Ising_model} is critical, otherwise we are in an
ordered or disordered phase.

The DOS and the localization measure \eqref{IPR} already give clear
indications of critical behavior.
\begin{figure}[ht!]
\includegraphics*[width=0.99\columnwidth]{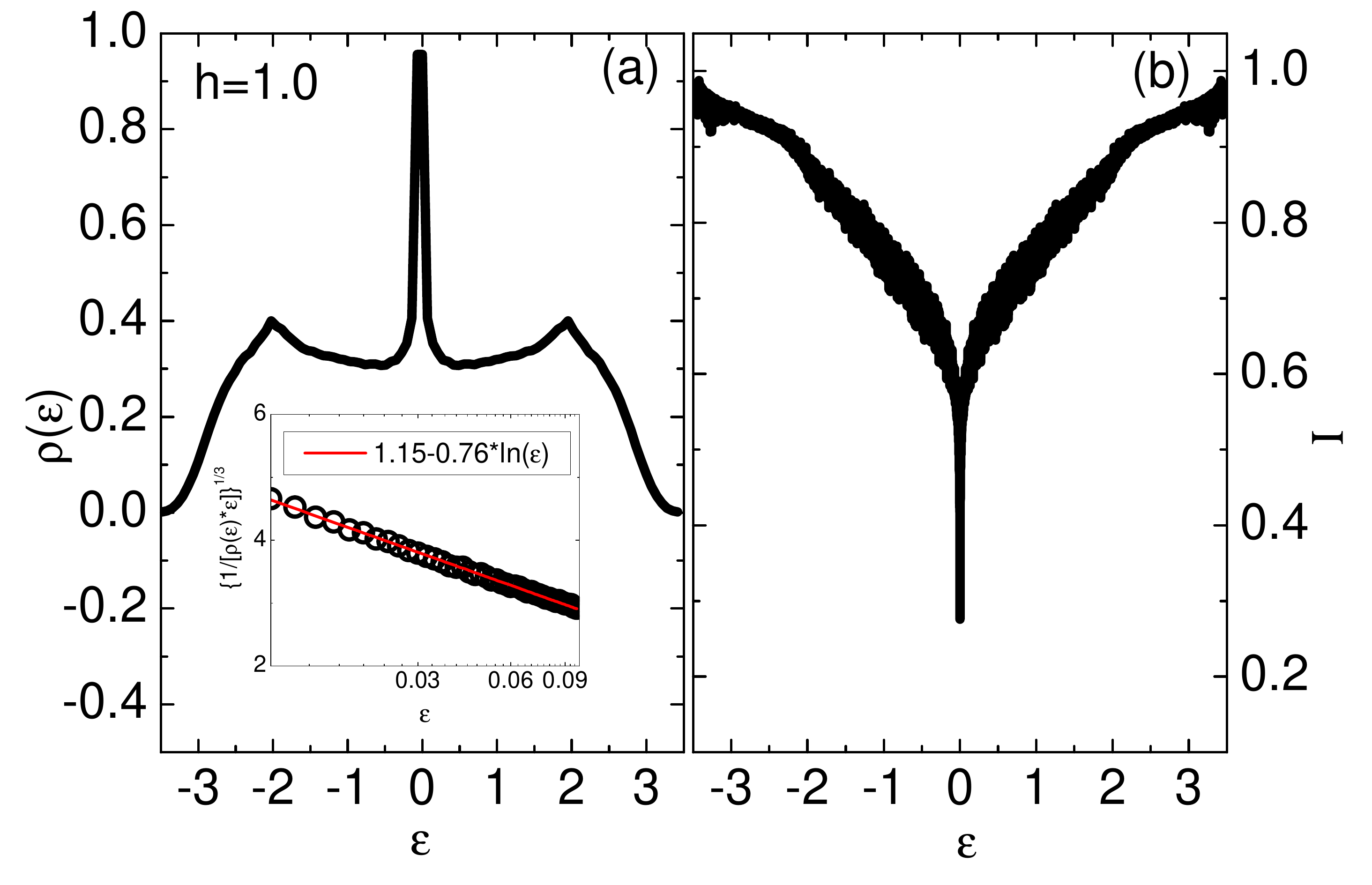}
\caption{(Color online) Critical random transverse Ising model with
  $J_i\in (0,1)$ and $h_i\in (0,1)$: (a) DOS as defined in
  Eq.~\eqref{DOS} using $10000$ samples with $100$ bins (main) and
  $3400$ bins (inset). (b) Localization measure \eqref{IPR} with
  $m=10$.}
\label{Fig18}
\end{figure}
As shown in Fig.~\ref{Fig18}(a) the density of states diverges in the
critical case for $|\varepsilon|\to 0$ in the thermodynamic limit. The
divergence is consistent with the analytically obtained formula
\eqref{DOS_XX} for the XX model. At the same time, the localization
measure $I_\varepsilon$ as defined in Eq.~\eqref{IPR} goes to zero for
$\varepsilon\to 0$ indicated a delocalization transition at zero
energy, see Fig.~\ref{Fig18}(b).

\begin{figure}[ht!]
\includegraphics*[width=0.99\columnwidth]{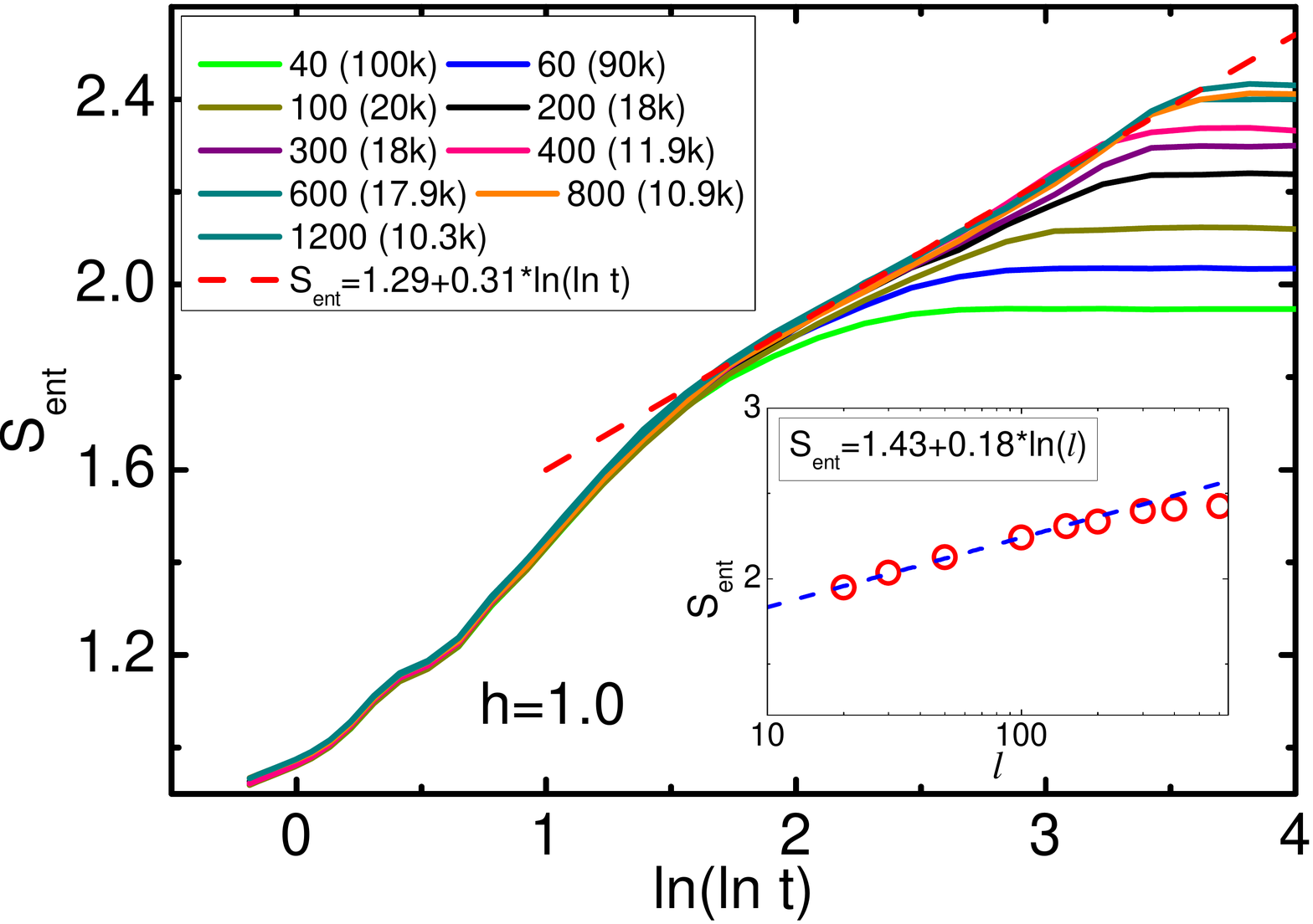}
\caption{(Color online) $S_{\rm ent}(t)$ in the random critical
  transverse Ising model with PBC for a quench from the initial state
  \eqref{Neel} with block size $\ell=N/2$. Inset: Saturation value
  $S_{\rm ent}(\ell,t\to\infty)$.}
\label{Fig20}
\end{figure}

The time evolution of the entanglement entropy in the random
transverse Ising model following a global quench has been studied
first by Igloi {\it et al.}, Ref.~\onlinecite{IgloiSzatmari}. For the
critical case they found a regime where $S_{\rm ent}(t)\approx
0.25\ln(\ln t) + \mbox{const}$. Furthermore, their data seem to
indicate that $S_{\rm ent}(\ell,t\to\infty)\approx
0.173\ln\ell+\mbox{const}$. This scaling would be qualitatively
consistent with the RSRG-X predictions, Eqs.~\eqref{Sent_XX} and
\eqref{Sent_XX2}, although the values of the prefactors as well as
their ratio is different. On the other hand, we have not found fully
conclusive evidence for a $\ln\ell$ scaling of the saturation value
for the XX model. To resolve this discrepancy we will, in the
following, extend the calculations in Ref.~\onlinecite{IgloiSzatmari}
to larger system sizes. Numerically, this is a demanding task. For
system sizes larger than $N=300$ we again encounter samples with very
small eigenvalues where the regular double precision diagonalization
routine fails to return a set of fully converged eigenvalues and
-vectors. Here the problem is even more severe because the linear
dimension of the Hamiltonian matrix is a factor $2$ larger ($2N\times
2N$ for the Ising model instead of $N\times N$ for the XX model). A
straightforward but very time consuming way to deal with this problem
is to diagonalize such problematic samples using multiprecision
routines. This is the way we have chosen to address this problem for
the XX model. Here we use a different approach where samples with
extremely small eigenvalues and a not fully converged spectrum are
simply discarded. This approach should not affect the data for times
$t<10^{14}$ but makes the results for larger times unreliable. A
further confirmation that this is a valid approach is that the
entanglement curves for various system sizes fall on top of each other
if $t<10^{14}$ and if the entanglement length $L(t)$ is smaller than
the block length $\ell$, see Fig.~\ref{Fig20}.

The entanglement entropy shows qualitatively the same behavior as for
the XX model with bond disorder and PBC. In particular, we find a
regime of $\ln(\ln t)$ scaling. The slope is comparable to the one we
found for the XX model with the bond couplings drawn from a box
distribution and overall consistent with previous results by Igloi
{\it et al.}, Ref.~\onlinecite{IgloiSzatmari}. Our main new result
concerns the saturation value $S_{\rm ent}(\ell,t\to\infty)$ shown in
the inset of Fig.~\ref{Fig20}: here we obtain data which, we believe,
are reliable at least up to $\ell=N/2=200$. In agreement with
Ref.~\onlinecite{IgloiSzatmari} where systems up to the maximum system
size $N=256$ were studied, we find that the saturation values are
consistent with a $S_{\rm ent}\sim\ln\ell$ scaling. The saturation for
chain lengths $N>400$, on the other hand, occurs at times $t>10^{14}$
and might already be affected by the discarding of samples with
extremely small eigenvalues. The ratio of the prefactors of the
$\ln\ln t$ and the $\ln\ell$ fits is $a/b=1.7$ in this case.

\begin{figure}[ht!]
  \includegraphics*[width=0.99\columnwidth]{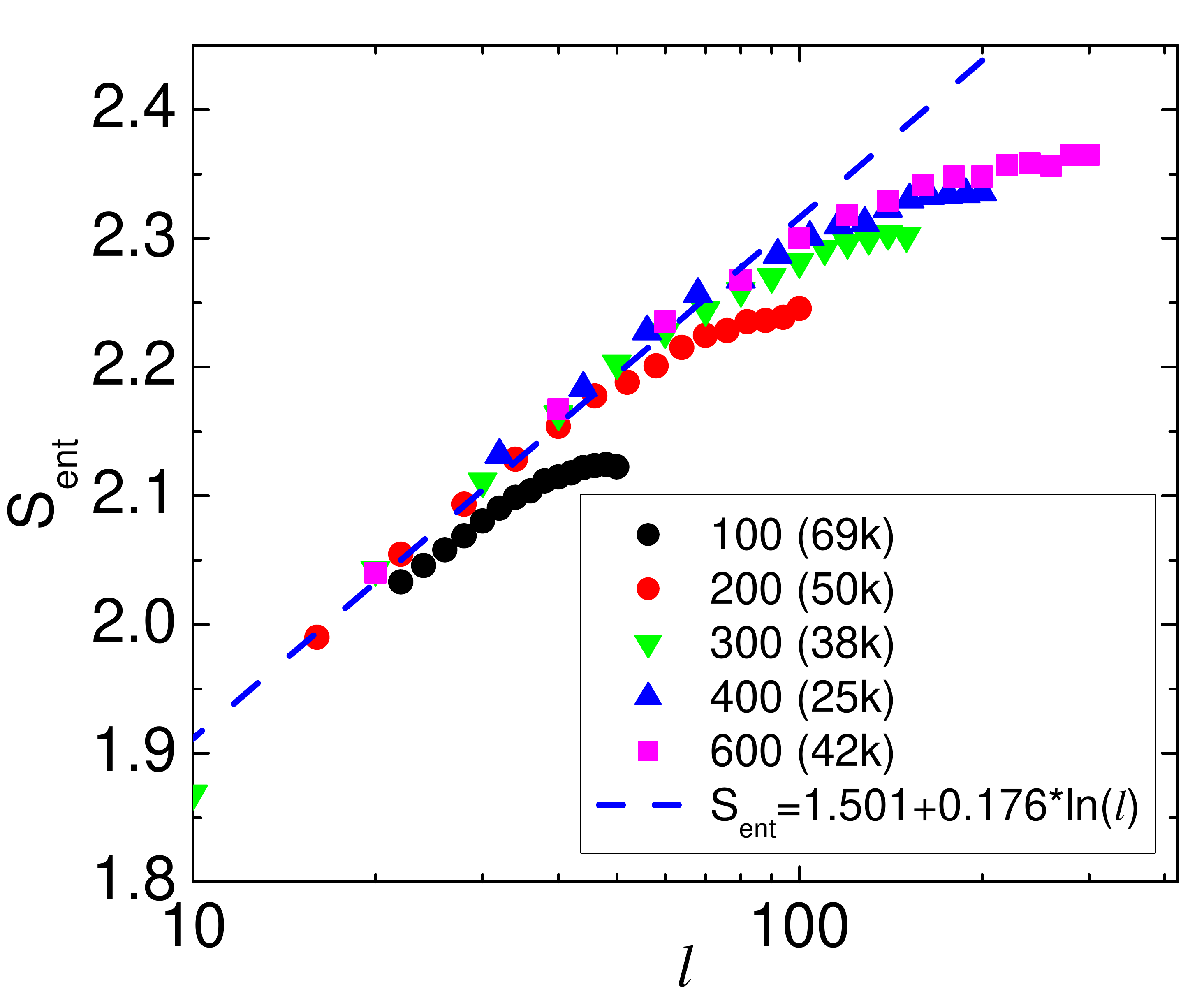}
  \caption{(Color online) Saturation value $S_{\rm
      ent}(\ell,t\to\infty)$ for a quench from the initial state
    \eqref{Neel} in the critical transverse Ising chain for various
    chain lengths $N$ as indicated. The dashed line is a fit
    representing potential logarithmic scaling.}
\label{Fig21}
\end{figure}

To investigate the scaling of the saturation value in more detail we
show data for different chain lengths $N$ and different block sizes
$\ell$ in Fig.~\ref{Fig21}.
These data cast some doubt if a $\ln\ell$ scaling really holds.
Deciding this question would require computationally very expensive
multiprecision calculations. As for the XX model we have to leave this
as an open problem for future investigations.

\section{Interacting models}
\label{Interaction}
We will concentrate on the XXZ chain with binary bond disorder
\begin{equation}
\label{XXZ}
H=\sum_i J_i\left\{ (c_i^\dagger c_{i+1} + h.c.)+\Delta (n_i-1/2)(n_{i+1}-1/2)\right\}
\end{equation}
and nearest-neighbor interaction $\Delta> 0$. Here $J_i=1\pm\delta$,
$n_i=c_i^\dagger c_i$ is the local density operator, and we are using
the fermionic representation of the XXZ model. The clean system,
$J_i\equiv J$, is critical for $-2\leq\Delta\leq 2$ and gapped
otherwise. The model \eqref{XXZ} has been investigated by RSRG-X for
the specific case of a quench from the density-wave state \eqref{Neel}
in Ref.~\onlinecite{VoskAltman} and by exact diagonalizations,
concentrating mainly on spectral properties, in
Ref.~\onlinecite{VasseurFriedman}.

\begin{figure}[ht!]
\includegraphics*[width=0.99\columnwidth]{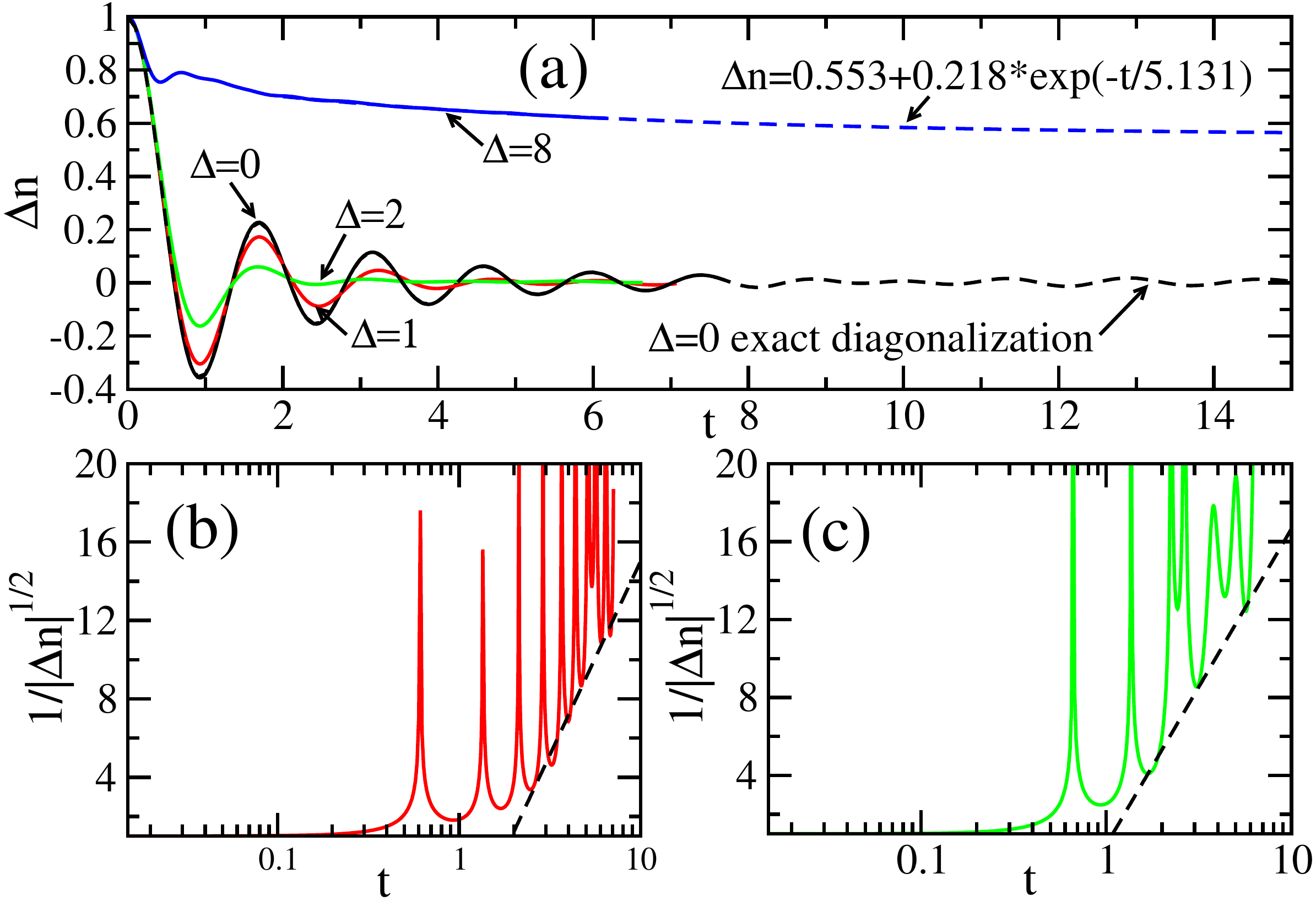}
\caption{(Color online) (a) $\Delta n(t)$ for binary disorder
  $\delta=0.2$ and different interaction strengths $\Delta$ obtained
  by LCRG for an infinite chain. The dashed line for $\Delta=0$ is the
  exact diagonalization result, the dashed line for $\Delta=8$ a fit.
  (b) $\Delta n(t)$ for $\Delta=1$ (solid line) and fit of the
  envelope (dashed line). (c) Same as in (b) but for $\Delta=2$.}
\label{Mstagg_int}
\end{figure}

Calculating the quench dynamics now becomes a true many-body problem
and the methods we have used before are no longer applicable. Instead,
we use exact diagonalizations of the many-body Hamiltonian for small
system sizes as well as the light-cone renormalization group
(LCRG),\cite{EnssSirker,AndraschkoEnssSirker} a variant of the DMRG,
to treat infinite XXZ chains. First, we consider the time evolution of
the order parameter $\Delta n(t)$. For the case without disorder,
numerical studies have found that the scaling of $\Delta n$ at short
and intermediate times seems consistent with an exponential
decay.\cite{BarmettlerPunk,BarmettlerPunk2,EnssSirker} Here we want to
study how this decay is affected by disorder. In Fig.~\ref{Mstagg_int}
we present data for binary disorder $\delta=0.2$ and different
interaction strengths $\Delta$.
For $\Delta=1$ and $\Delta=2$ the data indicate that $\Delta n(t)\to
0$ for $t\to\infty$. As shown in Fig.~\ref{Mstagg_int}(b) and (c),
respectively, the decay could possibly be consistent with
Eq.~\eqref{Sent_XX3}. Note, however, that in the non-interacting case
where data for much larger times are available (see
Fig.~\ref{Fig_bin_mstagg}) a clear scaling over several orders of
magnitude in time is not observed. What is clear, however, is that the
decay for $\Delta=1$ and $\Delta=2$ does not follow a simple
exponential as in the clean case, suggesting that the system is no
longer in the same ergodic phase. This becomes even more apparent for
$\Delta =8$ where the density wave order becomes effectively frozen
and a fit of the data seems consistent with $\Delta n\neq 0$ at
infinite times.

\begin{figure}[ht!]
\includegraphics*[width=0.99\columnwidth]{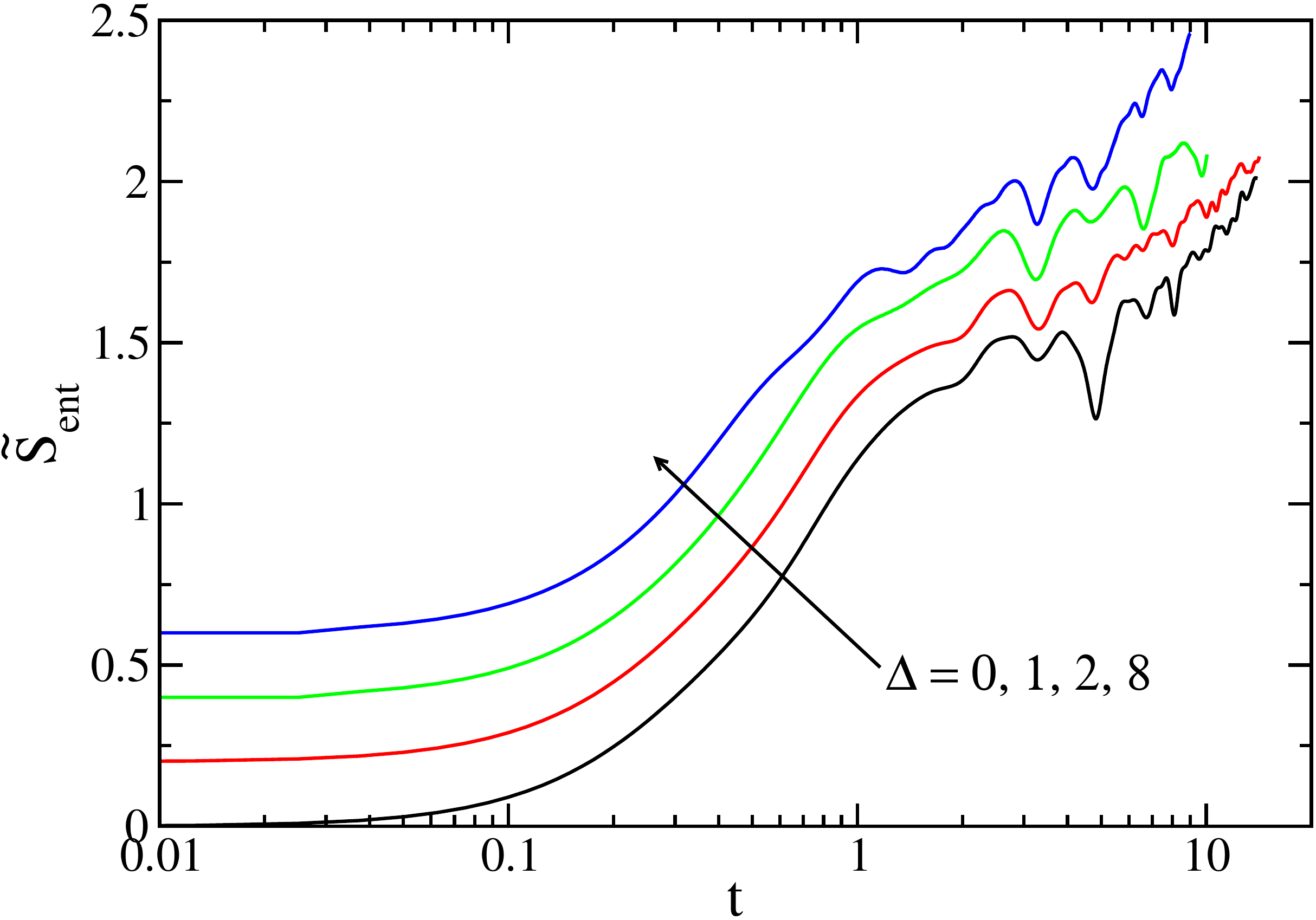}
\caption{(Color online) LCRG data for $\tilde S_{\rm ent}(t)$ of a
  system consisting of real sites and ancilla sites which realize
  binary bond disorder with $\delta=0.92$ exactly. Subsequent curves
  are shifted by $0.2$.}
\label{Fig_Stilde}
\end{figure}

Next, we consider the entanglement entropy in the interacting case.
In the LCRG algorithm the binary bond disorder is realized by ancilla
sites which are prepared in a completely mixed state. A partial trace
then yields the reduced density matrix of real {\it and} ancilla
sites.\cite{AndraschkoEnssSirker} The ancilla sites, however, are
static so that the entanglement entropy $\tilde S_{\rm ent}(t)$ of the
system with real and ancilla sites will show the same scaling behavior
as the entanglement entopy $S_{\rm ent}(t)$ of the system with real
sites only. The absolute values of the two entropies will, however, be
different. In Fig.~\ref{Fig_Stilde}, $\tilde S_{\rm ent}$ is shown for
strong binary disorder $\delta=0.92$ and different interaction
strengths.
After an initial increase, we find a regime where the entanglement
entropy seems to scale approximately logarithmically for all
interaction strengths including the noninteracting case and the
isotropic case $\Delta=2$. Theoretically, however, these cases are
expected to show quite different behaviors. For the quench from the
density-wave state and with $|\Delta|<2$ RSRG-X predicts
$S_{\rm{ent}}\sim (\ln t)^{2/\phi}$ with $\phi=(1\sqrt{5})/2$ being
the golden ratio. For the isotropic case arguments based on results
for SU(2)$_k$ chains suggest that in the Heisenberg limit
($k\to\infty$) the system becomes ergodic and shows volume law
entanglement scaling.\cite{VasseurPotter} For the time interval shown,
such differences are not observed. While LCRG data for longer times
can, in principle, be obtained, the calculations require a substantial
amount of computing time and it is unclear if a factor $2-3$ in time
would be sufficient to change this picture.

\begin{figure}[ht!]
\includegraphics*[width=0.99\columnwidth]{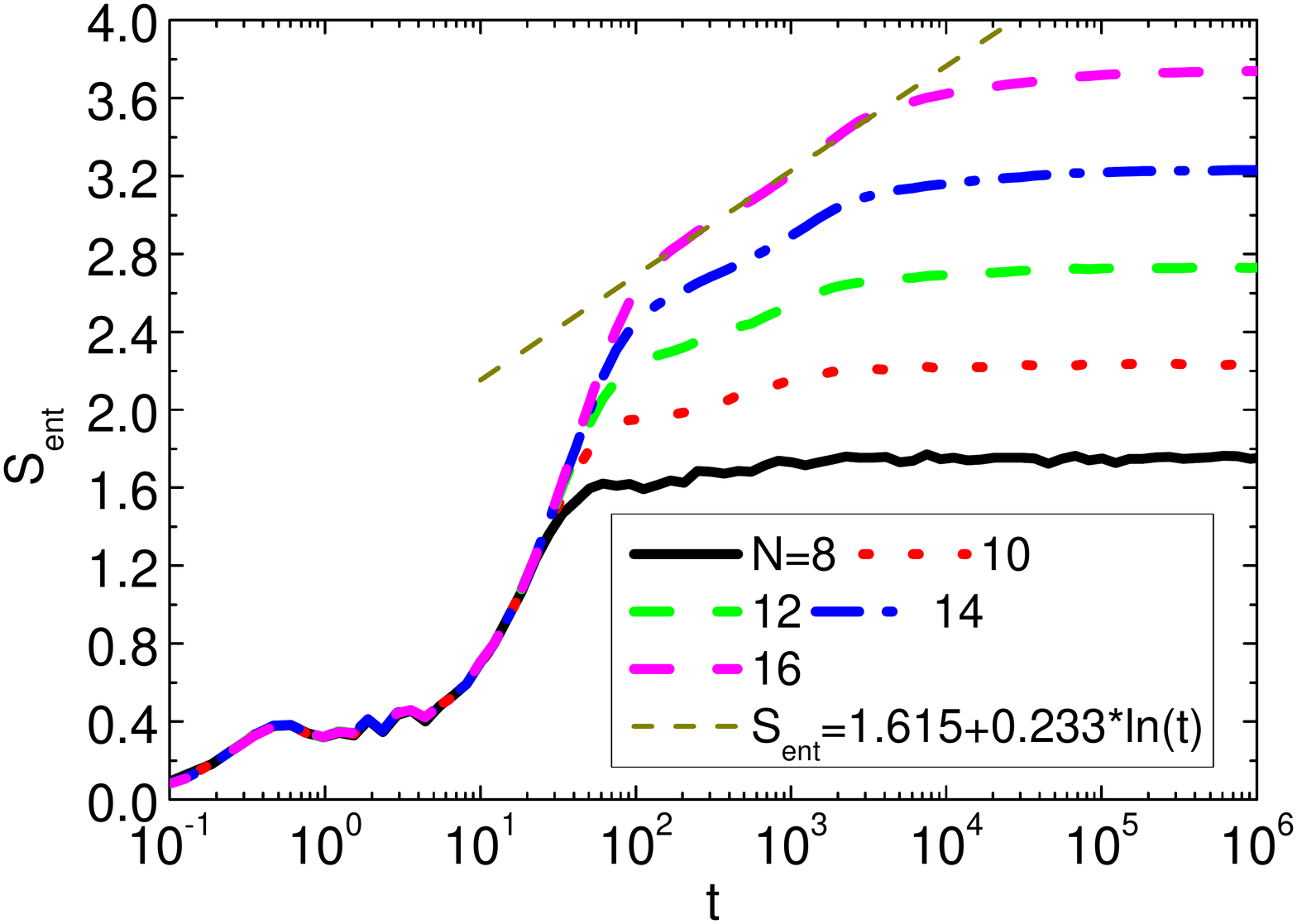}
\caption{(Color online) Exact diagonalization data for the isotropic XXZ model ($\Delta=2.0$)
  with binary bond disorder $\delta=0.92$. Averages are performed over
  $15000$ samples for $L<16$ and $11600$ samples for $L=16$.}
\label{Fig_HB_ED}
\end{figure}

Next, we turn to the exact diagonalization of small systems where long
times are accessible. We concentrate on the case $\Delta=2$ which
corresponds to the isotropic Heisenberg model if the Hamiltonian
\eqref{XXZ} is formulated in terms of spin-$1/2$ operators. In this
case we can make use of the $SU(2)$ symmetry which allows to
diagonalize larger systems than in the anisotropic case where $SU(2)$
is broken. In Fig.~\ref{Fig_HB_ED} data for several system sizes are
shown.
For $N\geq 12$, a logarithmic scaling seems to emerge at times which
are much larger than the simulation times in the LCRG calculations.
This is very different from our previous study of binary potential
disorder\cite{AndraschkoEnssSirker} where we could clearly distinguish
the Anderson insulator from the many-body localized system for times
accessible by LCRG. The results seem to indicate that there exists a
many-body localized phase for the bond-disordered Heisenberg
model. However, we cannot exclude that longer chains show a different
scaling consistent with thermalization as predicted in
Ref.~\onlinecite{VasseurPotter}.

\section{Conclusions}
\label{Conclusions}
We have numerically analyzed the entanglement growth in disordered
noninteracting and interacting quantum chains following a quantum
quench. While most papers so far have concentrated on the case of
potential disorder, our study has focussed on bond disordered models.
The latter systems are highly relevant theoretically because the real
space renormalization group (RSRG) can be applied in this case
allowing for a direct comparison between renormalization group
predictions and numerical data.

As a first step, we have compared the localization of the single
particle wavefunctions in the XX model in the potential disordered
with the bond disordered case. While potential disorder leads to a
localization length $\xi_{\rm loc}$ which only weakly depends on the
energy $\varepsilon$ of the wavefunction, a delocalization
transition---where the localization length diverges---occurs at
$\varepsilon=0$ for bond disorder. This has profound consequences for
the build-up of entanglement: $S_{\rm ent}(t)$ for the potential
disordered case shows an initial power-law increase followed by a
saturation. The $S_{\rm ent}(t)$ curve becomes independent of system
size $N$ and block size $\ell$ for $N,\ell\gg\xi_{\rm loc}$. For bond
disorder, on the other hand, the behavior is much more complex. An
initial approximately logarithmic increase, $S_{\rm ent}\propto
-\ln(1/t+\alpha)$ with $\alpha\ll 1$, is followed by a $S_{\rm
  ent}\sim\ln\ln t$ scaling over several orders of magnitude in time
and a saturation which depends on the block size.

The main focus of our work has been on a comparison of the numerical
results for $S_{\rm ent}(t)$ and for the density wave order parameter
$\Delta n(t)$ with predictions by the RSRG. For the XX model there are
three main predictions for a block of size $\ell$ in an infinite
system following a quench from a density wave state: (a) In the
scaling regime at long times $S_{\rm ent}\sim a\ln\ln t$, (b) the
saturation value scales as $S_{\rm ent}\sim b\ln\ell$, and (c) the
order parameter decays as $\Delta n\sim 1/\ln^2 t$. Numerically, we
find the decay of the order parameter to be consistent with the RSRG
result. The analysis of the entanglement entropy, on the other hand,
turns out to be much more complicated. Here extremely long times are
necessary to study saturation as a function of block size $\ell$. In
order to reach such times we have performed very time-consuming
multiprecision calculations which are needed because the spectrum of
bond disordered chains contains many extremely small eigenvalues. Even
with such data at hand, an unbiased numerical confirmation of the RSRG
predictions for the entanglement entropy turned out to be out of
reach. The main question which we have to leave open is as to whether
the saturation value scales as $S_{\rm ent}\sim b\ln\ell$ as predicted
by RSRG or saturates in the limit of infinite block size, $S_{\rm
ent}\sim -\ln(1/\ell+\alpha)$ where $0<\alpha\ll 1$ is a constant. For
the time dependence of the entanglement entropy this means that we
cannot decide if $S_{\rm ent}\sim a\ln\ln t$ is just a transient
behavior valid in a certain time range or if it does indeed hold for
all times in the limit $\ell\to\infty$. Furthermore, we find that even
if we take a $\ln\ell$ scaling as a given, a best fit yields a ratio
$a/b\approx 1.5$ which significantly deviates from $a/b=2$ expected
based on the RSRG analysis.

The RSRG scaling predictions should also be valid for the critical
random transverse Ising chain. This model can be mapped to free
fermions so that a numerical study based on single-particle wave
functions is possible. Our results for the entanglement entropy in
this case are qualitatively in agreement with the results for the XX
model. We again find $S_{\rm ent}\sim a\ln\ln t$ in a certain time range
but cannot obtain reliable data for large enough systems and long
enough times to decide whether or not the predicted $S_{\rm
  ent}\sim b\ln\ell$ scaling of the saturation value holds. Assuming
that the scaling holds, we find from a best fit that $a/b\approx 1.7$
which again deviates from the RSRG result.

Finally, we studied the interacting XXZ model with binary bond
disorder. Using an infinite-size density matrix renormalization group
algorithm we find that the entanglement entropy shows a $S_{\rm
ent}\sim\ln t$ scaling both in the interacting as well as in the free
fermion case $\Delta=0$. In the accessible time range it is thus not
possible to distinguish an Anderson insulator from a system which is
expected to be in a (critical) many-body localized phase. The exact
diagonalization of small interacting systems with large binary
potential disorder shows that the true scaling regime is only reached
at times which are much larger than the simulation times in the
density matrix renormalization group calculations. The exact
diagonalizations also seem to provide indications that the Heisenberg
model at strong binary bond disorder is in a many-body localized
phase. However, due to the limited system sizes which are accessible
numerically the theoretically predicted ergodicity of the
bond-disordered Heisenberg model---based on results for SU(2)$_k$
chains---cannot be excluded.

\acknowledgments We acknowledge support by the Natural Sciences and
Engineering Research Council (NSERC, Canada) and are grateful for the
computing resources and support provided by Compute Canada and
Westgrid. In particular, we would like to thank G.~Shamov for his help
in installing and using the MPACK libraries.




\end{document}